\documentclass[twoside]{article}

\usepackage{lipsum} 
\usepackage{graphicx}
\usepackage{caption}
\usepackage[sc]{mathpazo} 
\usepackage[T1]{fontenc} 
\usepackage{booktabs}
\usepackage{microtype} 
\usepackage[style=apa,sortcites=true,sorting=nyt,backend=biber]{biblatex}
\usepackage[margin=0.8in,columnsep=18pt]{geometry} 
\usepackage{multicol} 
\usepackage[labelfont=bf,textfont=it]{caption} 
\usepackage{float} 
\usepackage{hyperref} 

\usepackage{lettrine} 
\usepackage{paralist} 

\usepackage{abstract} 

\usepackage{titlesec} 
\usepackage[utf8]{inputenc}
\titleformat{\section}[block]{\bfseries\scshape\centering}{\thesection.}{1em}{} 
\titleformat{\subsection}[block]{\bfseries\centering}{\thesubsection.}{1em}{} 
\titleformat{\subsubsection}[block]{\bfseries\centering}{\thesubsubsection.}{1em}{}
\usepackage{fancyhdr} 
\title{\vspace{-15mm}\fontsize{16pt}{10pt}\selectfont\textbf{The UW Virtual Brain Project: An immersive approach to teaching functional neuroanatomy}} 

\author{
\normalsize
Karen B. Schloss$^{1,2}$\thanks{Correspondence concerning this article should be addressed to Karen Schloss, University of Wisconsin-Madison, 330 North Orchard Street, Room 3178
Madison, WI 53715. E-mail: kschloss@wisc.edu}, Melissa A. Schoenlein$^{1,2}$, Ross Tredinnick$^{2}$, Simon Smith$^{2}$,\\
\normalsize
Nathaniel Miller$^{1,3}$, Chris Racey$^{4}$, Christian Castro$^{5}$, and Bas Rokers$^{3}$\vspace{2mm}\\
\normalsize 
$^{1}$Department of Psychology, University of Wisconsin--Madison \\
\normalsize 
$^{2}$Wisconsin Institute for Discovery, University of Wisconsin--Madison\\
\normalsize              
$^{3}$Program in Psychology, NYU Abu Dhabi\\
\normalsize              
$^{4}$School of Psychology, University of Sussex\\
\normalsize 
$^{5}$Collaborative for Advancing Learning and Teaching, University of Wisconsin--Madison\\
\\
\normalsize
©2021, American Psychological Association. This paper is not the copy of record and may not\\
\normalsize
exactly replicate the final, authoritative version of the article. Please do not copy or cite\\
\normalsize
without author's permission. The final article will be available, upon publication, via its \\
\normalsize
DOI: 10.1037/tps0000281\\
\vspace{-8mm}
}

\date{}

\begin{document}

\maketitle 

\thispagestyle{fancy} 


\begin{abstract}

\noindent Learning functional neuroanatomy requires forming mental representations of 3D structure, but forming such representations from 2D textbook diagrams can be challenging. We address this challenge in the UW Virtual Brain Project by developing 3D narrated diagrams, which are interactive, guided tours through 3D models of perceptual systems. Lessons can be experienced in virtual realty (VR) or on a personal computer monitor (PC). We predicted participants would learn from lessons presented on both VR and PC devices (comparing pre-test/post-test scores), but that VR would be more effective for achieving both content-based learning outcomes (i.e test performance) and experience-based learning outcomes (i.e., reported enjoyment and ease of use). All participants received lessons about the visual system and auditory system, one in VR and one on a PC (order counterbalanced). We assessed content learning using a drawing/labeling task on paper (2D drawing) in Experiment 1 and a Looking Glass autostereoscopic display (3D drawing) in Experiment 2. In both experiments, we found that the UW Virtual Brain Project lessons were effective for teaching functional neuroanatomy, with no difference between devices. However, participants reported VR was more enjoyable and easier to use. We also evaluated the VR lessons in our Classroom Implementation during an undergraduate course on perception. Students reported that the VR lessons helped them make progress on course learning outcomes, especially for learning system pathways. They suggested lessons could be improved by adding more examples and providing more time to explore in VR.

\end{abstract}

\textbf{Public Significance Statement.} We designed and evaluated interactive 3D narrated diagrams to teach functional neuroanatomy. These lessons can be experienced on desktop PCs and in virtual reality (VR), and are helpful for teaching undergraduates about structure and function of perceptual systems in the human brain.

\begin{multicols}{2} 


To learn functional anatomy, such as how sensory information is processed in the human brain, students must form mental representations of 3D anatomical structures. Evidence suggests forming mental representations is easier for learners when they are presented with 3D models (i.e., different views can be rendered by translation and rotation), compared with 2D images (see \textcite{yammine2015} for a meta-analysis). This benefit of 3D models, at least in part, is because piecing together multiple views from 2D images incurs a cognitive load that detracts from learning the content, especially for learners with lower visual-spatial ability \parencite{bogomolova2020, cui2016}. 

Prior studies have suggested physical models are better than computer models for illustrating gross anatomy \parencite{preece2013, wainman2018, wainman2020, khot2013}. However, physical models are limited in their potential to illustrate dynamic, functional processes, such as how neural signals are triggered by sensory input and propagate through a perceptual system. Given that our focus is on functional anatomy, we will focus our discussion on computer-based models only.  

The present study is part of the UW Virtual Brain Project, in which we have developed and assessed a new approach for teaching students about functional anatomy of perceptual pathways. Previous computer-based 3D models of the human brain were geared toward teaching medical students about gross anatomy \parencite{adams2011, allen2016, ekstrand2018immersive, drapkin2015, Kockro2015, stepan2017, cui2017}.\footnote{Studies evaluating the efficacy of these 3D models used a myriad of comparison conditions that differed from the 3D models in multiple dimensions. Thus, it challenging to form general conclusions from their results (see \textcite{wainman2020} for a discussion of this issue).} In contrast, our lessons give learners guided, first-person view tours through ``3D narrated diagrams'' illustrating the functional anatomy of the human brain. We use the term ``3D narrated diagram'' to refer to 3D models combined with labels and verbal descriptions, analogous to content found in textbook diagrams with corresponding text. They can also include animations that illustrate dynamic aspects of the system. Thus, 3D models form the basis for the environment used to teach students about sensory input, system pathways, and system purposes, which are key learning outcomes in an undergraduate course on sensation and perception. 

Our aim was to develop structured, self-contained lessons for an undergraduate audience, which harnessed principles for effective multimedia learning \parencite{Mayer2009}. These principles have previously been shown to facilitate learning in a variety of domains. For example, using visual cues to signal students where to look during a lesson can help them learn about neural structures (signaling principle)  \parencite{jamet2008}. Learners benefit from having self-paced controls through a lesson, compared with experiencing system-paced continuous animation (segmenting principle) \parencite{hasler2007}. And, receiving input from multiple modalities (audio narration plus visual illustration) can be better than receiving visual input alone (modality principle) \parencite{harskamp2007}.

The UW Virtual Brain 3D narrated diagrams can be viewed on personal computer monitors (referred to as ``PC''; the same image is presented to both eyes) or in virtual reality using a head mounted display (HMD) with stereoscopic depth (referred to as ``VR''; different images are presented to each eye \footnote{We note earlier literature used the term ``VR'' in reference to viewing 3D models on 2D displays (e.g., computer monitors), rather than immersive head mounted displays (see \textcite{wainman2020} for a discussion of this issue). In this article, we reserve the term ``VR'' for head mounted displays, like an Oculus Rift, Oculus Go, or HTC Vive.}). In the VR version, the brain is room-sized, so learners can ``immerse'' their whole body inside the brain. This study investigated whether students made significant gains in content-based learning outcomes from the Virtual Brain lessons, and whether viewing device (VR vs. PC) influenced the degree to which learners achieved content-based and experience-based learning outcomes. Content-based learning outcomes included being able to describe (draw/label) key brain regions and pathways involved in processing visual and auditory input. Experience-based learning outcomes included finding the lessons enjoyable and easy to use.

We predicted that learners would make significant gains in content-based learning outcomes from lessons experienced in both VR and PC viewing (compared to a pre-test baseline), but VR viewing would be more effective. We also predicted VR would be more effective for achieving experience-based learning outcomes. Previous work strongly supports our prediction for experience-based learning outcomes, demonstrating that VR facilitates enjoyment, engagement, and motivation, compared with less immersive experiences \parencite{stepan2017, makransky2020, parong2018, pantelidis2010, hu2017}. However, prior evidence concerning our prediction that VR would better support content-based learning outcomes is  mixed. Research on learning 3D structure and spatial layout suggests VR should facilitate learning, but research on narrated lessons suggests VR may hinder learning, as discussed below. 

Research on learning 3D anatomical structure suggests stereoscopic viewing facilitates learning compared to monoscopic viewing of the same models, at least when viewing is interactive. A meta-analysis reported that viewing interactive stereoscopic 3D models provided a significant benefit, compared with viewing interactive monoscopic 3D models (i.e., the same image was presented to both eyes, or the image was presented to one eye only) \parencite{bogomolova2020}. For example, \textcite{wainman2020} found students learned better when stereoscopically viewing 3D models compared to when one eye was covered while using a VR HMD. The additional depth information provided by stereopsis likely contributes to these enhanced learning outcomes \parencite{bogomolova2020, wainman2020}. Evidence suggests that stereoscopic information is especially beneficial for 3D perception under interactive viewing conditions where head tracking-based motion parallax information and task feedback are available \parencite{fulvio2017}, perhaps because viewers tend to discount stereoscopic information under passive viewing conditions \parencite{fulvio2020}. This may explain why the contribution of stereopsis to achieve learning outcomes was more limited under passive viewing \parencite{alkhalili2014} and fixed viewpoint rendering \parencite{chen2012, luursema2008}. 

A separate line of studies testing the ability to remember spatial layout in new environments suggests that VR facilitates spatial memory. Comparisons between learning in VR or on a desktop PC suggest participants were better at navigating a virtual building \parencite{ruddle1999} and recalling the spatial location of objects \parencite{krokos2019} in VR. These results have been explained by increased presence (feelings of ``being there'') \parencite{sanchez2005} in VR, due to greater immersion  supported by proprioceptive and vestibular sensory information available during VR experiences \parencite{krokos2019, ruddle1999}. Increased presence enables learners to leverage spatial cues in the environment to facilitate memory (i.e., the method of loci or the memory palace technique) \parencite{krokos2019}.

Although work on stereopsis and spatial memory suggests VR will help with learning spatial structure in the Virtual Brain lessons, research comparing narrated lessons viewed in VR or on PCs suggests VR might hinder learning. Studies on narrated lessons about scientific procedures (e.g., DNA sample preparation) reported no difference \parencite{ makransky2020} or worse performance for VR \parencite{makransky2019, makransky2020} compared to PC viewing. One exception in which VR facilitated learning was when it was paired with enactment using physical objects in-between the VR lesson and testing \parencite{makransky2020}. The general concern about narrated VR lessons is that presence from immersion in VR can distract learners from the content in the lesson, which impedes learning \parencite{makransky2019, makransky2020, parong2018}. Thus, it is possible that adding additional cognitive load from narration will diminish or even reverse the benefits of learning 3D structure from added stereopsis and increased presence in VR.

In the following sections, we first describe the general methods for designing and implementing the UW Virtual Brain Project lessons. We then present the results of two laboratory experiments that assessed learning outcomes of lessons presented in VR vs. PC viewing (Experiment 1 and Experiment 2). Finally, we discuss how we implemented the lessons in an undergraduate course on the psychology of perception and present results from student evaluations (Classroom Implementation).


\section*{General Methods}

\begin{figure*}
  \includegraphics[width=\linewidth]{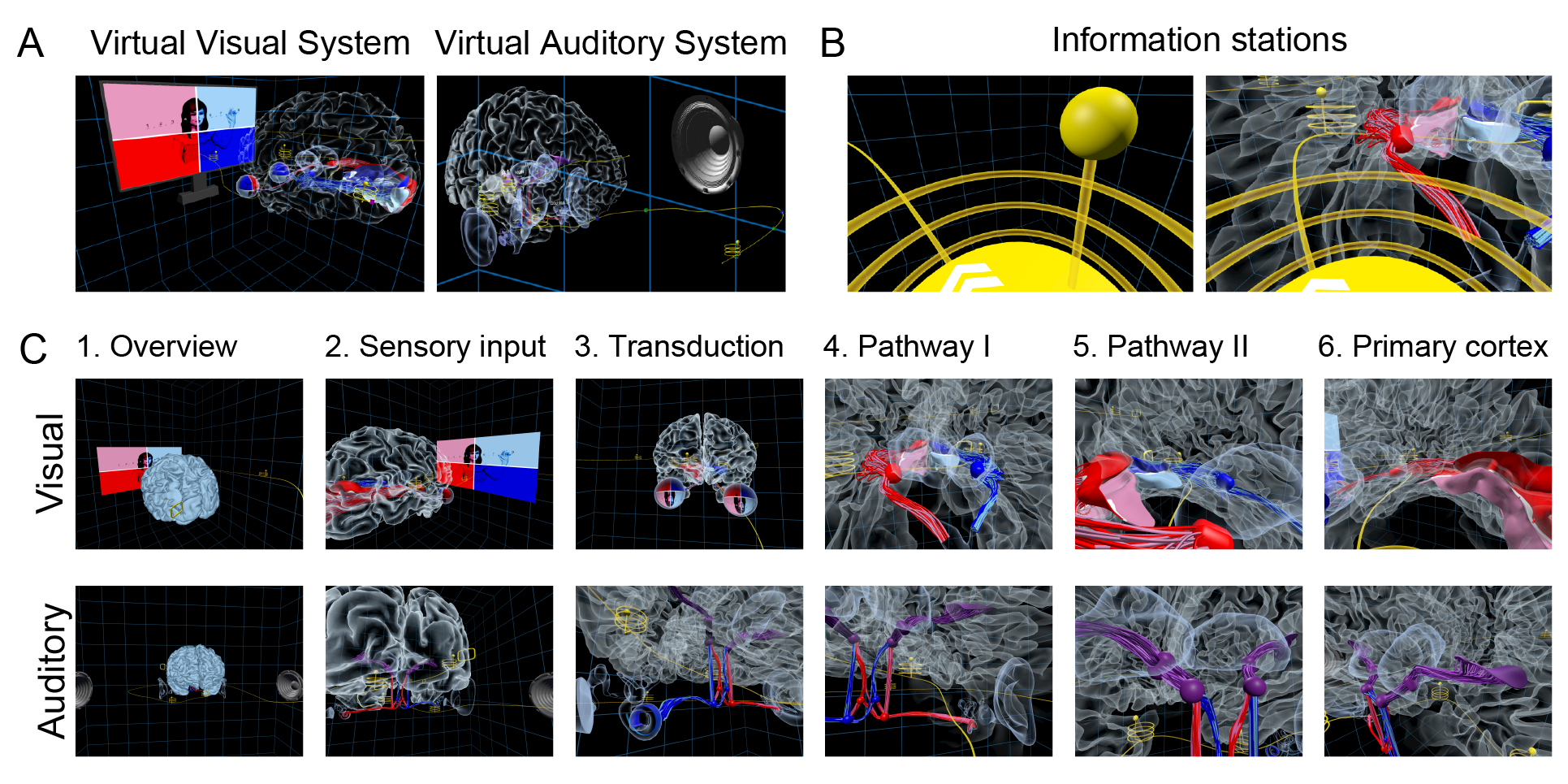}
  \caption{Illustration of the of the UW Virtual Brain Project. (A) Screenshots of the Virtual Visual System and Virtual Auditory System. (B) Information stations, where participants learn about the structures and pathways via audio narration. (C) Each lesson has 6 stations that learners visit along a track at key points in the perceptual system pathway.}
  \label{fig:SystemPlatform}
\end{figure*}

We created and evaluated two UW Virtual Brain lessons, the \textit{Virtual Visual System} and \textit{Virtual Auditory System}. Learners could travel along a track and stop at information stations to hear narration about key structures and pathways involved in perceptual processing (Figure \ref{fig:SystemPlatform}). The interactive diagrams in our Virtual Brain lessons were adapted from figures in a popular sensation and perception textbook \parencite{wolfe2014}, and constructed from human neuroimaging data.

The UW Virtual Brain Project lessons leveraged several principles for creating effective materials for multimedia learning \textcite{Mayer2009}. In our lessons, verbal narration was combined with visual input (multimedia principle) in a way that enabled learners to listen and look at the same time (modality principle). Through the narration, we provided verbal cues about where to look (e.g., ``look to your left, you will see...'') to help learners align the visual and verbal input (signaling principle). The lessons were self-paced (segmentation principle), as learners controlled their motion and triggered narration by choosing when to enter an information station. We avoided additional elements unrelated to the content (coherence principle) by only including neural structures and labels that were relevant to the perceptual system featured in a given lesson. 

For any kind of VR-based visualization, motion sickness can lead to considerable discomfort \parencite{stanney2020identifying}. We employed several design considerations to mitigate motion sickness. First, we optimized our code to maintain the maximum frame-rate of the VR headset. Second, the participant was in control of all movement, eliminating the possibility of any drastic and unexpected motion signals. Third, the trajectory along which the participant could move was always visible, allowing the participant to anticipate the visual consequences of any initiated movement. We evaluated the efficacy of these design considerations using the Simulator Sickness Questionnaire (SSQ) \parencite{kennedy1993simulator} and a questionnaire assessing participants' enjoyment and ease of use in the different conditions, which we describe in the Measures section of Exp. 1A. The SSQ results are presented in the Supplemental Material.

Figure \ref{fig:DesignPipeline} outlines our pipeline for constructing the virtual environments. We began with a T1-weighted anatomical MRI scan and used FreeSurfer \parencite{fischl2012freesurfer} to extract the pial surface of the brain. This approach was similar to \textcite{ekstrand2018immersive}. We generated cortical regions of interest by extracting surfaces from FreeSurfer's default segmentation and cortical surface generation and Glasser's Human Connectome Project Multimodal Parcellation Atlas \parencite{glasser2016multi}. For some subcortical structures, we estimated their location based on gross anatomy and rendered them manually. We generated major white matter pathways using probabilistic tractography and manually recreated smaller pathways. We then imported all geometry into the Unity game engine and added features including voice-over playback, text rendering, and navigation. Additional details on the history of VR brain rendering can be found in the Supplemental Material.

The \textit{Virtual Visual System} and \textit{Virtual Auditory System} (Figure \ref{fig:SystemPlatform}A) each have six information stations, which start outside of the brain (Station 1) and follow along a track from sensory input (Station 2), to transduction (Station 3), to midbrain regions and pathways (Stations 4 and 5), to primary cortex (Station 6) as shown in Figure \ref{fig:SystemPlatform}C. When learners arrive at a station, glowing yellow rings appear around the perimeter of the station (Figure \ref{fig:SystemPlatform}B) and voice-over narration provides information relevant to that location. After the audio finishes, the rings fade away, and learners can continue along the track. The locations and narration for the two perceptual systems were as parallel as possible and the lessons were equal in length (\textasciitilde5 minutes). See the Supplemental Material for descriptions of the lesson experiences and narration scripts. The lessons can be downloaded from https://github.com/SchlossVRL/UW-Virtual-Brain-Project. 

\begin{figure*}
  \includegraphics[width=\linewidth]{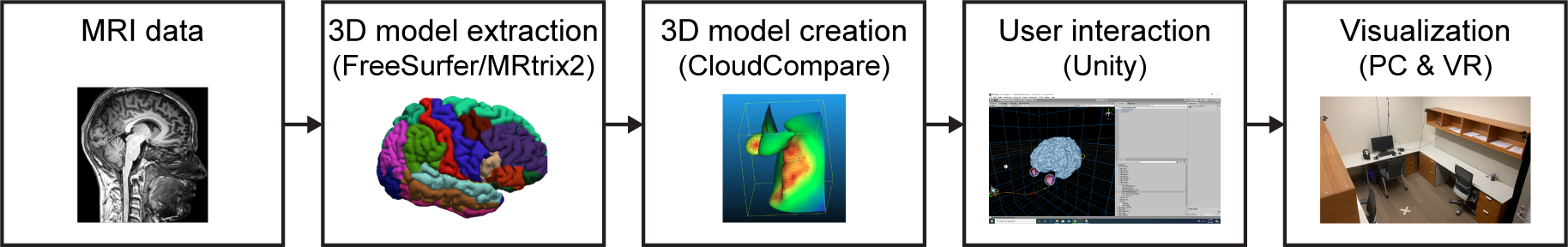}
  \caption{Pipeline for creating the UW Virtual Brain lessons, beginning with an MRI scan and ending with lessons that can be used on PC desktops or in VR.}
  \label{fig:DesignPipeline}
\end{figure*}

In Experiments 1 and 2, the VR and PC setups were powered by a Dell Alienware Aurora workstation with a nVidia GeForce 970 GPU. In the VR version we used an Oculus Rift CV1 with 360 degree tracking enabled. Three Oculus camera sensors were placed in the upper corners of a 6-ft $\times$ 8-ft tracked space. Participants stood in the middle of the space, and the Oculus cord was suspended from above, enabling full 360-degree body rotation. Although our participants stood, the VR lessons could be done while seated on a swivel chair. Participants heard narration through built in speakers in the HMD and interacted with the environment using Oculus Touch controllers. They moved forward/backward along the track by pressing the Oculus Touch joystick and looked around the room by moving their head. The HMD tracked this head movement (6-degrees of freedom head tracking) and updated the image to match the head motion. In the PC version learners sat in front of a 24-inch Samsung monitor and heard narration through headphones. They used the left/right mouse buttons to move forward/backward along the track and slid the mouse in any direction to ``look around'' (rotate the view). In the Classroom Implementation we used wireless Oculus Go HMDs which support 3-degrees of freedom head tracking (rotation only) and the Oculus Go controller to move along the track.

For each device in the experiments, participants completed a practice lesson that introduced them to the virtual environment and the device controls. The environment included a model of the brain's cortical surface (no structures/pathways). Audio narration instructed participants about how to use the controls for the given device and asked them to practice moving forward/backward and looking around. In the VR practice, participants also used a virtually rendered eye chart to adjust the interpupillary distance of the Oculus lenses to minimize blurriness.

\section*{Experiment 1}
Experiment 1 compared the effects of PC vs. VR viewing on achieving content-based and experienced-based learning outcomes. Assessments of content-based learning outcomes were done by drawing/labeling on paper. Exp. 1A and 1B were the same except Exp. 1B had twice the sample size to increase statistical power.

\section*{Experiment 1A} 


\subsection*{Methods}

\subsubsection*{Participants} 
60 undergraduates (30 female, 29 male, 1 no report; mean age = 19.1) participated for credit in Introductory Psychology at University of Wisconsin--Madison. Data from three others were excluded due to experimenter error. A power analysis estimating a medium effect ($d = .5$) for a two-tailed paired t-test comparing PC vs. VR ($\alpha = .05$, power of .9) suggested a sample of $n = 44$, but we rounded to $n = 60$ to be more conservative. All participants had typical color vision (screened using H.R.R. Pseudoisochromatic Plates \parencite{hardy2002hrr}) and typical stereo vision (screened using the RANDOT\textregistered Stereovision test). For all experiments in this study, all participants gave informed consent and the UW--Madison IRB approved the protocol.

\subsubsection*{Design and Procedure}  
Figure \ref{fig:ExpProcedureMethods} shows an overview of the experiment procedure (Figure \ref{fig:ExpProcedureMethods}A), the experiment design (Figure \ref{fig:ExpProcedureMethods}B), and the testing materials (Figure \ref{fig:ExpProcedureMethods}C). The design included 2 devices (VR and PC; within-subjects) $\times$ 2 device orders (VR-first or PC-first, between-subjects) $\times$ 2 perceptual system-device pairings (visual-VR/auditory-PC or auditory-VR/visual-PC; between subjects) ($n = 15$/group, randomly assigned).

During the consenting process, we emphasized that participants could end the experiment at any time if they felt unwell. During the experiment, participants first completed the SSQ as a baseline measure for motion sickness symptoms. Second, they completed the first lesson block, which included a paper pre-test, a practice experience for the given device, the lesson, and a post-test. Third, participants completed the second SSQ to assess how they were feeling after the first lesson. Fourth, participants completed the second lesson block, which was the same as the first, except using a different device (e.g., if VR was used in the first lesson block, PC was used in the second lesson block) and different perceptual system (e.g., if the visual system was used in the first lesson block, the auditory system was used in the second lesson block). Fifth, participants completed another SSQ to assess how they were feeling after the second block. Last, participants completed the experience questionnaire. The procedure lasted approximately 1 hour. The time to complete the VR condition was about 3 minutes longer than the PC condition due to extra time adjusting the HMD in the practice room before the VR lesson.

\subsubsection*{Measures}
 
\textbf{Content learning.} We assessed content learning using a pre-test/post-test design for each perceptual system with the same test provided at both time points. We used a drawing/labeling task, which aligned with the content-based learning outcomes of describing key regions and pathways of the perceptual systems. We used the same pre-test and post-test because it was a controlled way of testing exactly what gains in knowledge were made during the lessons. A limitation of this approach is that learners were primed on the critical information prior to starting the lesson. However since the design was consistent across the PC and VR conditions, this priming could not account for differences between conditions.

\begin{figure*}[ht!]
  \includegraphics[width=\linewidth]{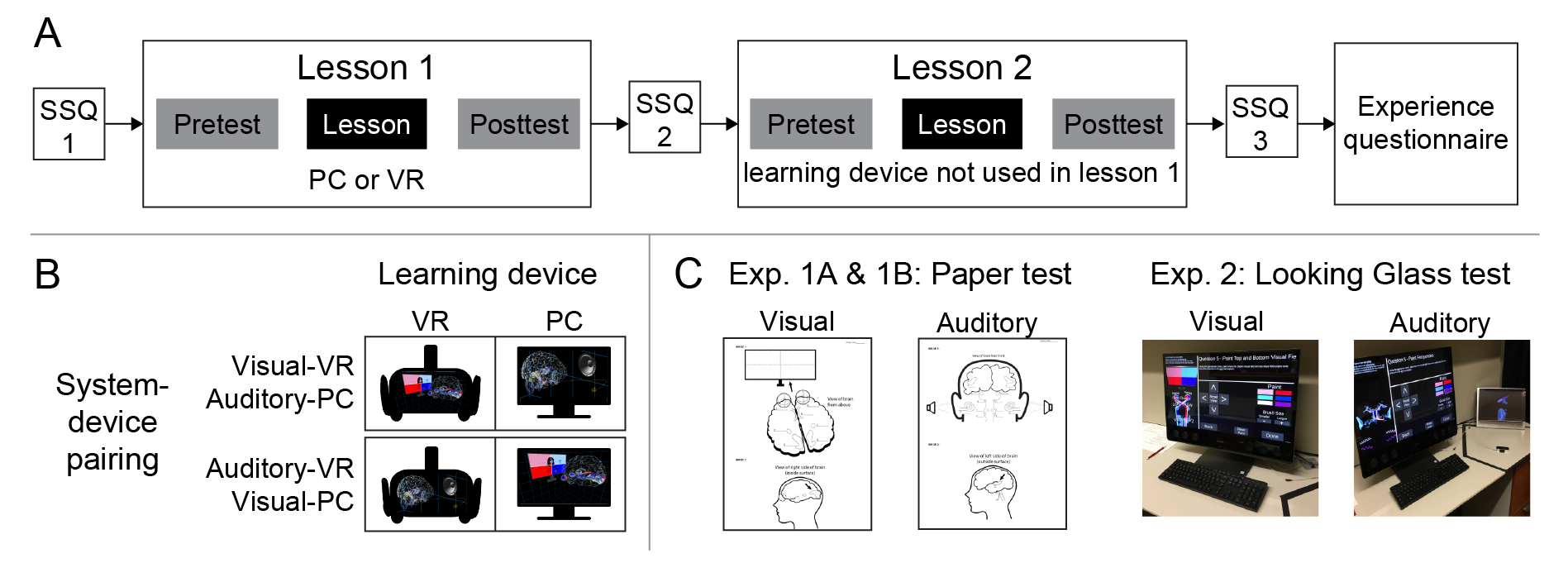}
  \caption{Overview of the experimental procedure. (A) In both Experiments 1 and 2 each lesson block featured a pre-test, the lesson (either desktop personal computer (PC) or virtual reality (VR)), and a post-test. Three simulator sickness questionnaires (SSQs) were administered throughout the course of the experiment, with an Experience questionnaire at the end. (B) Illustration of the experimental design. We randomized the 2 learning device (within-subjects) $\times$ and 2 perceptual system-device pairings (between-subjects). Each row represents the lessons experienced by a participant. Not explicitly represented is device order. If the figure presented represents VR-first, the PC-first groups would mirror that (between-subjects). (C) Illustration of Paper tests (Exp. 1A and 1B) and Looking Glass tests (Exp. 2) used to assess content-based learning outcomes for each perceptual system lesson. See Supplemental Material for full tests and instructions and for larger images of the Looking Glass display (Figure C10).}
  \label{fig:ExpProcedureMethods}
\end{figure*}

Tests for both perceptual systems were matched in number and types of questions. Tests for each system included two line drawing images of the brain from different perspectives (Figure \ref{fig:ExpProcedureMethods}C), which resembled perspectives experienced during the lessons. Participants responded to five questions by coloring/labeling structures and drawing pathways on the two images (image 1 for questions 1-4; image 2 for question 5), using pens in three hues (red, blue, purple), with two pens per hue (light/dark), and one black pen (7 pens total). The questions built on one another, requiring participants to use the hues/lightness of the pens to indicate how the sensory input from the world travels into and through the perceptual system and projects onto the system's structures. Participants were encouraged to answer all questions, even if guessing. If they made a mistake, they were told to ask for a new blank page. See Supplemental Material for full tests and instructions (Figures C1-C5).

In the visual system test, the first test image consisted of a TV screen and an axial slice of the brain featuring two eyes and outlines of the optic chiasm, LGNs, and V1 structures. The second test image showed a sagittal slice of the brain, featuring the right hemisphere, with the corresponding eye, LGN, and V1. Participants were told to use the colored pens as follows: ``Reds: information processed from the left visual field; Blues: information processed from the right visual field; Black: Labeling structures.'' The questions are summarized as follows: (a) Use the colored pens to indicate (color in) the four parts of the visual field in the TV screen (i.e., ``dark red pen: bottom left visual field), (b) Color in the quadrants on the eyes, with respect to where the visual field projects (i.e., ``dark red pen: quadrant(s) of the eyes where the bottom left visual field projects''), (c) Use the black pen to write the names of the structures on the blank lines, (d) Use the ``appropriate colors'' (based on previous responses) to draw the path from both the right and left visual field through all relevant structures, (e) On image 2, indicate (color) where the bottom and top visual field project on the marked structure (which was V1). 

In the auditory system test, the first test image featured a coronal slice of a head with ears and a brain, flanked by speakers ``producing'' sound waves. In the brain, the cochleas, MGNs, and A1 structures were outlined and circles represented the cochlear nuclei, superior olives, and inferior colliculi. The second test image showed a profile view of the left side of the head and had outlines of MGN and A1. Participants were told to use the colored pens as follows: ``Reds: information processed by the ear seen on your right; Blues: information processed by the ear seen on your left; Purples: information processed by both ears; Black: Labeling structures.'' The questions for the auditory test paralleled the visual except in place of the visual field (TV), participants colored in the sound waves from the speakers and drew the pathways in regard to low and high frequency sounds. Also, instead of the retina and V1, participants colored the parts of the cochlea and A1 in reference to processing lower/higher frequencies.

\textbf{Experience Questionnaire.} We assessed experience-based learning outcomes by having participants rate seven items on a Likert scale from 1 (``Not at all'') to 7 (``Very Much''). The items asked how much participants (a) found the experience awe inspiring, (b) found the experience aesthetically pleasing, (c) enjoyed the experience, (d would like to use this kind of experience for their own studies about the brain in the future, (e) would recommend the experience to a friend for learning about the brain, (f) would recommend the experience to a friend to do for fun, (g) found ease with using the control system to move around and see what they wanted to see. This task was done on paper. 

\textbf{Simulator Sickness Questionnaire (SSQ)}. We assessed a subset of symptoms from the SSQ \parencite{kennedy1993simulator}. For each symptom (headache, nausea, eye strain, and dizziness with eyes open) participants indicated how they felt by circling  ``None'', ``Slight'', ``Moderate'', or ``Severe'' (scored as 1-4). The SSQ results are reported in the Supplemental Material, see Figure D1. For all experiments, mean responses to all four symptoms were between none and slight at all time points, and no participant reported severe symptoms. This task was done on paper.    

\subsection*{Results and Discussion}
\subsubsection*{Scoring} Each test was scored by two independent raters using an 18-item rubric with each item worth 1 point (see the Supplemental Material). Prior to collecting and scoring the data, we collected data from five pilot subjects to fine-tune the rubric. The two raters used an initial rubric to independently score the pilot data. They then discussed discrepancies in their scores and updated the rubric to make it more precise. No changes to the rubric were made once testing in Exp. 1A began. 

During the experiment, each test page was given a random code that represented subject number and condition. The raters did not have access to these codes, so they could not identify which tests were pre-tests versus post-tests, which were from VR or PC lessons, and which belonged to the same participant. To evaluate inter-rater reliability, we correlated each of the raters' scores over 240 tests [60 participants $\times$ 2 lessons (visual, auditory) $\times$ 2 tests (pre-test, post-test)]. The correlation was high (\textit{r} = .94), so we averaged the scores across raters for each participant. We then calculated change in performance for each participant as the post-test scores minus pre-test scores (ranging from -18 to 18) for each device. Data for all experiments can be found at https://github.com/SchlossVRL/UW-Virtual-Brain-Project.

\subsubsection*{Content questions}
\begin{figure*}[ht!]
  \includegraphics[width=\linewidth]{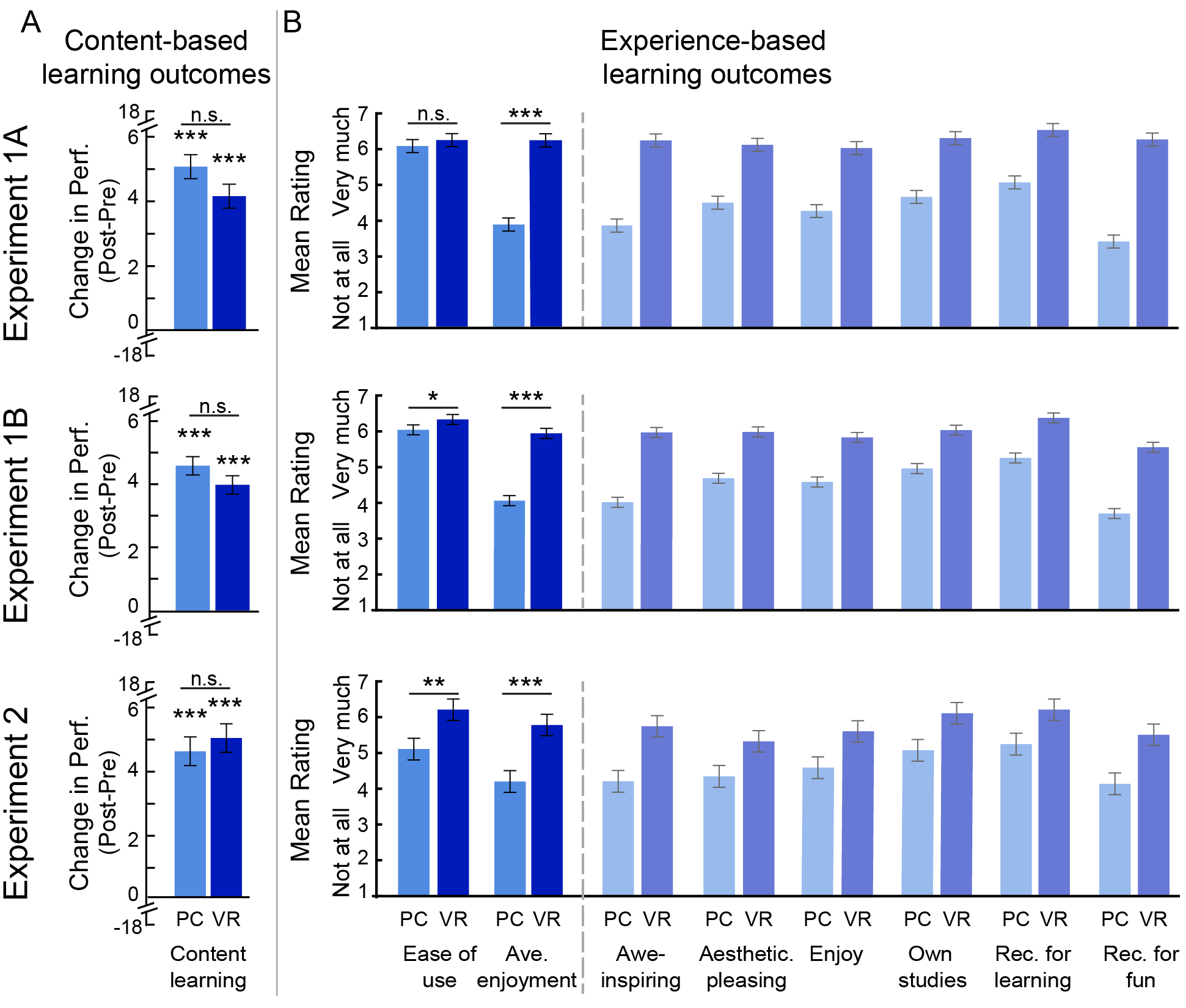}
  \caption{Results of learning outcome assessments as a function of device (PC vs. VR). (A) Mean change in test performance and (B) mean ratings for ease of use and ``average enjoyment'' for personal computer (PC; light bars) and virtual reality (VR; dark bars) devices, averaged over participants, testing order, and perceptual system in Experiments 1 and 2. The data to the right of the vertical dashed line in (B) correspond to the 6 individual items used to calculate ``average enjoyment''. Error bars represent standard errors of the means. $*p<.05$, $**p<.01$, $***p<.001$}
  \label{fig:ExperienceSTEM}
\end{figure*}

Figure \ref{fig:ExperienceSTEM}A shows mean change in test performance for the PC and VR devices, averaged over participants, testing order, and perceptual system-device pairing. t-tests against zero showed that change scores were significantly positive for both PC and VR ($t(59) = 13.05, p < .001, d = 1.68$, and $t(59) = 11.69, p < .001, d = 1.51$, respectively), indicating that participants learned from the Virtual Brain lessons on both devices. A paired samples t-test did not reveal a significant difference in learning between devices ($t(59) = 1.59, p = .118, d = .21$). 

In an exploratory analysis to test for effects of device order and perceptual system-device pairing, and their interaction with device (mixed-design ANOVA: 2 devices (VR vs. PC; within-subjects) $\times$ 2 device orders (VR-first vs. PC-first; between-subjects) $\times$ 2 perceptual system-device pairing (visual-VR/auditory-PC vs. auditory-VR/visual-PC; between-subjects)), there was no effect of device ($F(1,56) = 2.47, p = .122, \eta_{p}^{2} = .041$), or system-device pairing ($F<1$), but there was a main effect of device order ($F(1,56) = 4.56, \textit{p} = .037, \eta_{p}^{2} = .075)$, in which the mean change in performance was greater for participants in the VR-first group than the PC-first group. The 2- and 3-way interactions were not significant ($Fs < 1.19$).

We also explored whether there was a difference in the pre-test performance for the two perceptual systems and found greater performance for the visual system ($t(59) = 7.00, p < .001, d = .90$). Given the way we coded the counterbalanced factors, an interaction between device and perceptual system-device pairing would indicate differences in change in performance for the visual and auditory systems. We do not see such a difference here, indicating learning was similar across perceptual systems.

\subsubsection*{Experience questionnaire} Figure \ref{fig:ExperienceSTEM}B shows the mean ratings for each of the seven experience questionnaire items. From visual inspection, VR ratings were much higher than PC for every item except ease of use. This suggests participants found the VR experience more awe-inspiring, enjoyable, aesthetically pleasing, and were more likely to use it themselves and recommend it to others for learning or for fun than the PC experience.  

Given that many of the items were highly correlated (see Table D1 in the Supplemental Material), we used Principle Components Analysis (PCA, with oblique rotation) to reduce the dimensions before conducting statistical tests to compare across devices.\footnote{Five of the participants left one item blank, so we inserted the mean of all other participants' responses for that item.} All items except `ease of use' loaded strongly onto the first principal component, which can be summarized as ``enjoyment'' (capturing 66\% of the variance). Thus, we used the mean of these items for subsequent analysis and refer to it as ``average enjoyment.'' Item 7, ease of use, was an outlier in the PCA, so we treated it separately. Paired samples t-tests comparing devices showed that average enjoyment was significantly greater for VR than for PC lessons ($t(59) = 9.50, p < .001, d = 1.23$), and there was no significant difference for ease of use ($t(59) = .86, p = .39, d = .11$). 
 
In summary, this experiment demonstrated that participants learned key content such as the brain regions and pathways involved in processing sensory information from the lessons experienced on both devices. There was no significant difference in content learning between devices, but assessments of experience-based learning outcomes showed that participants enjoyed the VR lesson significantly more than the PC lesson. There was no significant difference in ease of use. 

In the \textit{a priori} power analysis for this experiment, we estimated a medium effect size comparing VR and PC devices for measures of content-learning, but a power analysis using our observed effect size ($\eta_{p}^{2} = .041$, when accounting for device order and perceptual system/device pairing) with power of .80, $\alpha = .05$ suggested we needed a larger sample size (112 participants) to observe an effect. We note that if this effect were significant, it would be in the opposite direction of our prediction (i.e., greater learning for PC than VR). To test this possibility, we conducted Exp. 1B as a direct replication of Exp. 1A with an increased sample size (120 participants to be more conservative).

\section*{Experiment 1B}

\subsection*{Methods}
The methods were the same as Exp. 1A except we increased the sample size based on the power analysis reported in Exp. 1A. 120 undergraduates (79 females, 36 males, 5 no report; mean age = 18.60, two no report) participated for extra credit in Introductory Psychology at UW--Madison. All had typical color and stereo vision and gave informed consent. Eight additional participants were excluded due to experimenter error or technical difficulties (4 participants), atypical color vision (1 participant), or atypical stereo vision (3 participants). 

\subsection*{Results and Discussion}
\subsubsection*{Content questions} The two raters from Exp. 1A scored the tests (inter-rater reliability: $r = .91$). As in Exp. 1A, change scores were significantly greater than zero for PC ($t(119) = 15.46, p < .001, d = 1.41$) and VR ($t(119) = 13.94, p < .001, d = 1.27$), indicating that participants learned on both devices (Figure \ref{fig:ExperienceSTEM}A). The t-test comparing devices averaged over all other factors again showed no significant difference between devices ($t(119) = 1.62, p = .109, d = .15$). Likewise, the full ANOVA [2 devices (VR vs. PC; within-subjects) $\times$ device orders (VR-first vs. PC-first; between-subjects) $\times$ 2 perceptual system-device pairings (visual-VR/auditory-PC vs. auditory-VR/visual-PC; between-subjects)] again showed no significant effect of device, even after having increased power ($F(1,116) = 2.95, p = .089, \eta_{p}^{2} = .025$). The other main effects were also not significant: device order ($F(1,116) = 2.01, p = .159, \eta_{p}^{2} = .017$), system-device pairing ($F<1$). There was a significant device $\times$ system-device pairing ($F(1,116) = 15.93, p < .001, \eta_{p}^{2} = .121$), which can be reduced to better performance for the visual system than the auditory system. That is because participants with the system-device pairing of visual-VR/auditory-PC had higher scores for VR (visual) compared to PC (auditory), whereas participants with auditory-VR/visual-PC pairings had higher scores for PC (visual) than VR (auditory) indicating overall greater learning for the visual system. This is further supported by a paired samples t-test comparing change in performance scores for visual and auditory lessons ($t(119) = 3.95, p < .001, d = .36$).
The other 2-way interactions and 3-way interaction were not significant: device $\times$ device order ($F<1$), perceptual system-device pairing $\times$ device order ($F<1$), 3-way interaction  ($F(1,116) = 2.30, p = .132, \eta_{p}^{2} = .019$). Examining the pre-test scores, the visual system scores were again significantly greater than the auditory system scores ($t(119) = 7.66, p < .001, d = .70$).

\subsubsection*{Experience questionnaire} As in Exp. 1A, we conducted statistical tests on mean enjoyment (averaging over six items) and ease of use. This data set includes only 114 out of 120 participants because six participants did not complete the survey. Mean enjoyment was significantly greater for VR than for PC ($t(113) = 9.16, p < .001, d = .86$) (Figure \ref{fig:ExperienceSTEM}B), as in Exp. 1A. Ease of use was also significantly greater for VR than PC ($t(113) = 2.39, p = .02, d= .22$), which was not significant in Exp. 1A (likely due to the smaller sample size in Exp. 1A).

In summary, Exp. 1B showed that with greater statistical power, there was no difference between VR and PC viewing on achieving content-based learning outcomes (learning occurred on both devices). However, increasing power may be responsible for the finding that participants rated VR as significantly easier to use than PC, which was only marginal in Exp. 1A.

\section*{EXPERIMENT 2}
Exp. 1 showed no differences between PC and VR devices for content learning. However, it is possible that a difference could still exist and our paper test measure was not sufficient to detect it.  Although paper assessments (2D drawing) may be the norm in the classroom, they may be limited in their ability to assess students' mental representations of 3D structures. Moreover, paper assessments were better aligned with 2D viewing on the PC than 3D viewing in VR. Thus, in Exp. 2 we examined whether testing under 3D viewing would reveal differences in learning from VR vs. PC devices. By comparing these results to Exp. 1, we could test for effects of alignment between learning and testing method (similar to \textcite{wainman2018}).

The most straightforward way to implement testing in 3D viewing would have been to test in the same VR device used for learning. However, in Exp. 1, testing was implemented using a different format (paper/pens) from the two devices used for learning (VR and PC), so we also wanted to use a different format for testing in Exp. 2. Thus, we used a Looking Glass 3D autostereoscopic display, which allowed 3D viewing without glasses via parallax barriers and multiple layers of displays. Participants interacted with the Looking Glass using a Leap Motion hand tracking controller, enabling them to complete analogous tests as in Exp. 1, but using their hands to draw/label in 3D.

\subsection*{Methods}
\subsubsection*{Participants}
48 undergraduates (29 females, 18 males, 1 no report; mean age = 19.17) participated for extra credit in Introductory Psychology at UW--Madison.\footnote{We planned to collect data for 120 participants to match Exp. 1B, but data collection was suspended due to COVID-19. The pattern of results of this experiment parallel those of Exp. 1, suggesting that the change in testing method does not change the main results, even with the reduced sample size.} All had typical color and stereo vision. Additional participants were excluded due to experimenter error (4 participants), not finishing in the allotted time (1 participant, reported eye strain), atypical color vision (2 participants), atypical stereo vision (1 participant).

\subsubsection*{Design, Displays, and Procedure}
The design, displays, and procedure were the same as in Exp. 1, except we replaced the paper drawing/labeling tests with analogous tests using a 15.6 inch Looking Glass autostereoscopic display system \parencite{dodgson2005autostereoscopic}, see Figure \ref{fig:ExpProcedureMethods}C. The overall display resolution was 3840 $\times$ 2160 px, with 45 separate views rendered at 768 $\times$  240 pixels/view. Participants interacted with the Looking Glass using a Leap Motion hand tracking controller and a 28-in touch screen PC (A Dell XPS 27-7760 All-in-One Desktop). Due to the novelty of the Looking Glass and the need to train participants on how to use it, the experiment typically lasted 15-30 minutes longer than Experiment 1 (approximately 1.5 hours).

The Looking Glass displayed the 3D model of the virtual brain from the lessons, except that it was resized to fit on the display. It featured different subsections of the brain, providing views similar to the outlines on the paper drawing/labeling tests (see Figures C8 and C9 in the Supplemental Material for images of the touch screen and corresponding views on the Looking Glass for each test question). 

The touchscreen contained four main sections: the questions (top), response buttons (right), a screenshot of the previously completed questions (left), and controls for changing the viewpoint of the display (middle). The questions were the same as the paper tests, except we replaced the use of pens with the response buttons (i.e., color swatches, labels). The questions were presented one at a time, with different response buttons activated to match the question (i.e., structure labels would replace the color swatches for labeling questions). Screenshots of completed questions appeared on left of the touchscreen, allowing participants to view their previous answers. Each test had four tasks analogous to the paper test: filling in structures, labeling structures, drawing pathways, and painting on structures (See Supplemental Material for details). For each task, participants used one hand to make selections on the touchscreen, and the other hand to complete the drawing task. The Leap Motion tracked their drawing hand and replicated its motion using a 3D hand model in the Looking Glass. Because the Looking Glass tasks were novel, participants received training on how to do the tasks prior to completing the first pre-test. They learned how to fill, draw, paint, and label parts of a 3D model house. Additional details on this training, including instructions (Figures C6 and C7) and examples displays (Figure C10), can be found in the Supplemental Material.

\subsection*{Results and Discussion}
\subsubsection*{Content questions} Pre- and post-tests were automatically scored, except for items requiring painting portions of the system (i.e., painting the halves of V1 to represent where the upper/lower visual field maps onto V1; four questions in the visual system and seven for the auditory system) and one question addressing fiber cross-overs in the visual system. These questions were scored by the same two raters from Experiment 1 for all 192 tests (48 participants $\times$ 2 tests $\times$ 2 experiences) following a rubric adapted from that of Exp. 1. Inter-rater reliability was high (\textit{r} = .98) and scores were averaged over raters for each participant.  

As in Exp. 1, change in performance was significantly positive for both PC ($t(47) = 9.55, p < .001, d = 1.38$) and VR ($t(47) = 12.08, p < .001, d = 1.74$), indicating participants learned using both devices (Figure \ref{fig:ExperienceSTEM}A). The t-test comparing devices averaged over all other factors again showed no significant difference between devices  ($t(47) = -.80, p = .428,  = .12$). Similarly, the full ANOVA [2 devices (VR vs. PC; within-subjects) $\times$ device orders (VR-first vs. PC-first; between-subjects) $\times$ 2 perceptual system-device pairing (visual-VR/auditory-PC vs. auditory-VR/visual-PC; between-subjects)] showed no significant effect of device ($F<1$). None of the other main effects or interactions were significant: device order ($F(1,44) = 2.87, p = .097, \eta_{p}^{2} = .061$), system-device pairing ($F<1$), device $\times$ system-device pairing ($F<1$), device $\times$ device order ($F(1,44) = 2.46, p = .124, \eta_{p}^{2} = .053$), system-device pairing $\times$ device order ($F<1$), 3-way interaction ($F<1$). Examining just the pre-test scores again indicates that the visual system scores were significantly greater than the auditory system pre-tests ($t(47) = 5.77, p < .001, d = .83$).

We next examined whether testing format (2D drawing on paper in Experiment 1 vs. 3D drawing on the Looking Glass in Experiment 2) resulted in different patterns of performance from PC or VR learning. We used a mixed design ANOVA with 2 lesson devices (PC vs. VR; within-subjects) $\times$ 2 testing devices (Paper vs. Looking Glass; between-subjects). There were no significant effects of lesson device ($F<1$) or testing device ($F(1,226) = 1.30, p = .254, \eta_{p}^{2} = .006$), and no interaction ($F(1,226) = 2.86, p = .092, \eta_{p}^{2} = .013$). Thus, we saw no significant effect of whether the testing format (2D vs. 3D drawing) was aligned with the learning format (2D vs. 3D viewing). However, it is noteworthy that the lack of effects related to testing device suggest participants could demonstrate their knowledge similarly using the novel Looking Glass as with familiar paper/pen testing.

\subsubsection*{Experience questionnaire} As in Exp. 1, average enjoyment was significantly greater for VR than PC ($t(28) = 4.08, p < .001, d = .76$) (Figure \ref{fig:ExperienceSTEM}B). As in Exp 1B, ease of use was also significantly greater for VR lessons than PC lessons ($t(28) = 3.32, p = .002, d = .62$).\footnote{The experience questionnaire data set includes 29/48 participants because several participants did not realize the items were split over two screens and thus did not complete the task, and the experimenters unfortunately did not notice until later.} 

In summary, Exp. 2 replicated Exp. 1, even though testing was conducted in a different format (i.e., drawing in 3D using stereoscopic depth rather than drawing on paper). Thus, aligning the assessment to 2D vs 3D viewing had no significant effect on performance.

\section*{Classroom Implementation}
Given that we developed these lessons to assist undergraduate students learning functional neuroanatomy of the brain, we implemented and evaluated our lessons in an undergraduate course,  \textit{Psychology of Perception} (UW--Madison, Spring 2019). Our goals were to (a) gather indirect measures of learning, which reflected students' self-perception of the efficacy of the Virtual Brain lessons for learning the material and (b) obtain feedback on which aspects of the lessons contributed to their learning and what aspects could be more useful.

With 25 Oculus Go VR headsets, approximately 80 students completed each 5-minute lesson within 20 minutes. The lessons were used at different times in the semester when the material in the Virtual Brain lessons was typically covered. The Virtual Visual System was used in Week 2 of the course, in place of slides typically used to cover that material. Students completed the VR lesson at the start of the class period, and the instructor immediately transitioned to lecturing on new material. The instructor observed that starting a class meeting with VR felt abrupt, and transitioning from the VR lesson to new material felt disjointed. Therefore, we revised our approach for the auditory system, implemented in Week 11. We embedded the VR lesson in a pre-VR lecture and post-VR activity, allowing for gradually-released instruction \parencite{bransford2000people} (i.e., I do it, I do it with you, you do it alone). In the 'I do it' phase, the instructor first lectured briefly on the material that would be covered in VR. In the `I do it with you phase', students completed the guided tour in VR. In the `you do it alone phase', students worked in pairs to enact the auditory system: one student ``held'' an imaginary brain and the other used their fingers to draw the pathways and point out the structures, which were listed on a slide to remind students where to ``stop'' on their imaginary tour. Next, one pair volunteered to demonstrate their imaginary tour, which transitioned focus back to the front of the classroom and fostered peer learning. Finally, the instructor moved on to new material. This approach also leveraged the pre-training principle for effective multimedia learning, by which people learn more deeply from multimedia lessons when they know the names and characteristics of main concepts before beginning the lesson \parencite{Mayer2009}.
 
At the end of the semester (Week 14), we conducted a voluntary survey in class to evaluate how the VR lessons contributed to achieving course learning outcomes. Students were informed that this study was separate from their course work and the instructor would not know who participated (a different member of the study team administered the survey).

\subsection*{Methods}
Students were told participation in the evaluation was voluntary and not part of their coursework ($n=53$ chose to participate). Participants were given a survey that first asked them to circle ``yes'' or ``no'' to indicate if they experienced each of the VR lessons used in class several weeks prior. If participants answered yes, they were asked to rate how much they thought the VR lesson helped advance their progress on three learning outcomes: (a) Describe the key brain regions involved in processing visual/auditory information and the pathways that connect them (\textit{System Pathways}), (b) Explain how sensory input from the world stimulates the visual/auditory system (\textit{Sensory Input}), and (c) Describe how the way information that is processed in the visual/auditory system supports the general purposes of the system (\textit{System Purpose}). Although all three learning outcomes were relevant to the material, one was most specific to the VR lesson (Systems Pathways). We were interested in whether students' progress ratings on this outcome would be greater than the other outcomes.

Participants responded using a 5-point scale by selecting: ``No apparent progress'' , ``Slight progress'', "Moderate progress'', ``Substantial Progress'', or ``Exceptional progress'' (coded as 1-5). There were also two free-response questions:  (1) What aspects of the Virtual Brain activities contributed to your learning? and (2) What would make the Virtual Brain activities more useful for your learning?

 \subsection*{Results and Discussion}
Two coders independently recorded the ratings and transcribed the free-response answers. Discrepancies were resolved by a third coder. The data can be found at https://github.com/SchlossVRL/UW-Virtual-Brain-Project.

\subsubsection*{Learning outcome ratings} 
\begin{figure*}[ht!]
  \includegraphics[width=\linewidth]{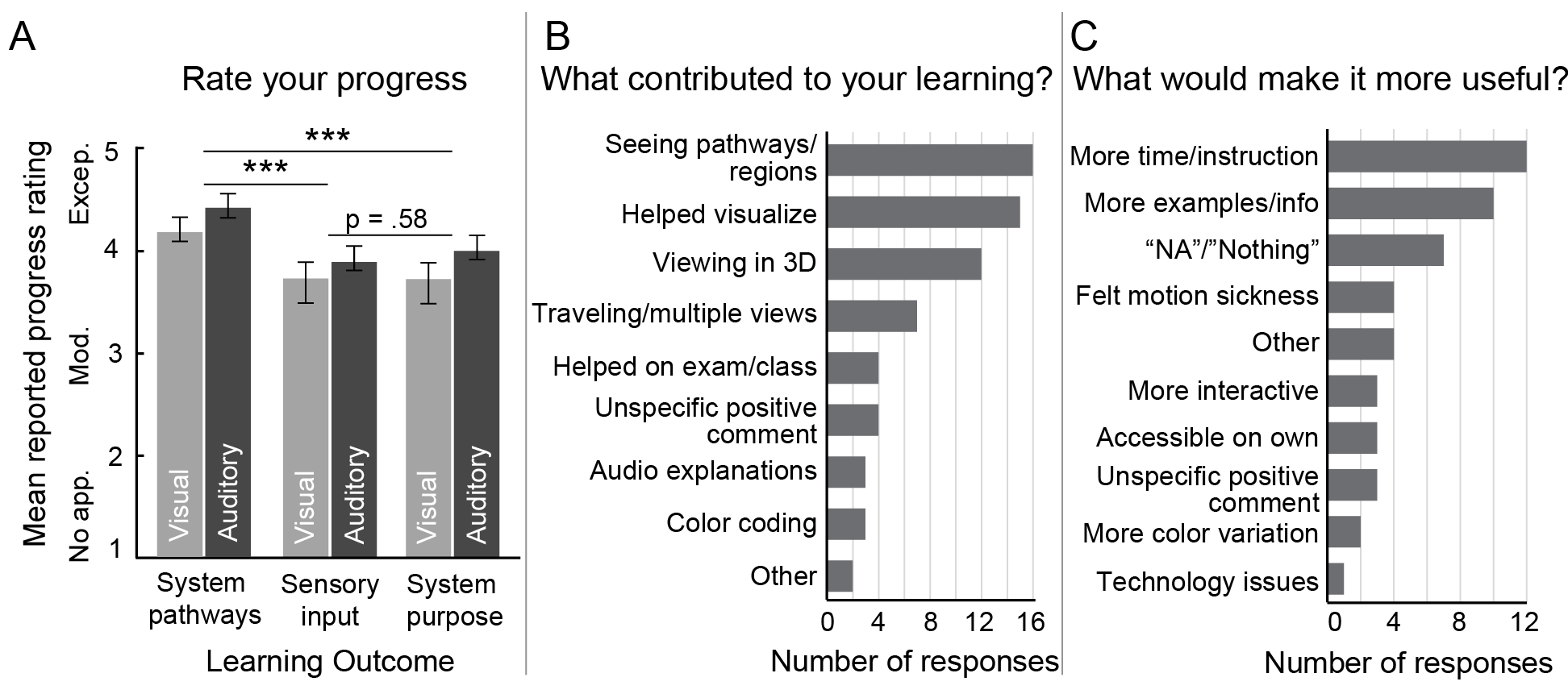}
  \caption{Student evaluations of the Virtual Brain lessons following the classroom implementation. (A) Mean self-report ratings for how the Virtual Visual System (light gray) and Auditory System (dark gray) affected progress on three learning outcomes, ranging from no apparent progress (no app.) to moderate progress (mod.) to exceptional progress (except.). Error bars represent standard errors of the means. (B) Frequency of free-responses sorted into each theme relating to what aspects of the Virtual Brain contributed to learning. (C) Frequency of free-responses sorted into each theme relating to what would make the activities more useful for learning. ``Other'' includes comments not directly related to the Virtual Brain and comments that were unclear. $***p<.001$}
  \label{fig:ClassImplement}
\end{figure*}

We analyzed data from participants that experienced both lessons ($n = 40$) so we could directly compare the visual and auditory systems. On average, students reported that VR lessons helped them make moderate to substantial progress on all learning outcomes (Figure \ref{fig:ClassImplement}A). A repeated-measures ANOVA comparing 2 perceptual systems (visual vs. auditory) $\times$ 3 learning outcomes (sensory input vs. system purpose vs. system pathways) revealed main effects of learning outcome ($F(1,78) = 16.14, p < .001, \eta_{p}^{2} = .293$) and perceptual system ($F(1,39) = 10.70, p = .002, \eta_{p}^{2} = .215$), with no interaction ($F<1$). The main effect of system indicated students reported more progress from the auditory system lesson than the visual system lesson. This may have been because the auditory lesson was embedded in pre-VR/post-VR activities, using gradual release of instruction \parencite{bransford2000people}, whereas the visual system was not. However, given that the auditory system lesson was used later in the semester, we cannot rule out alternative explanations based on timing. By the time students experienced the Virtual Auditory System, they may have developed greater familiarity/comfort with VR and greater general knowledge about perception. Further work is needed to disentangle these possibilities. 

We also compared learning outcomes using pairwise t-tests (Holm-corrected). Ratings were higher for system pathways, the key learning outcome for Virtual Brain lessons, compared with sensory input ($t(78) = 5.05, p < .001, d = .80$) and system purpose ($t(78) = 4.42, p <.001, d = .70$). There was no significant difference between sensory input and system purpose ($t(78) = -.57, p = .57, d = -.09$). Given that the learning outcome factor did not interact with sensory system (reported above), we can conclude that the higher ratings for system pathways carried through both lessons, and was not significantly affected by whether the lesson was done earlier or later in the semester.  

The results of this assessment are limited in that they are based on indirect student self-reports rather than external, direct measures of learning gains. It is possible that students' reports could be attributed to the novelty of VR in the classroom rather than the learning per se. However, novelty alone can not explain the differences seen between perceptual systems and between learning outcomes, especially because the lesson that was rated higher (auditory system) was completed after learners had already experienced VR earlier in the semester and was therefore less novel.

\subsubsection*{Free-response feedback} To organize the free responses, one coder reviewed the transcribed responses and identified themes for each question based on common repeating or related key words. She and two additional coders then completed a cursory thematic analysis by sorting responses into those themes. In order for a response to be counted for a theme, at least two coders had to be in agreement. Figures \ref{fig:ClassImplement}B and \ref{fig:ClassImplement}C show the frequency of responses across themes for each question. We focus our discussion on themes that emerged among at least one quarter of the responses (the top two themes for each question in Figures \ref{fig:ClassImplement}B-C). 

In response to what aspects of the Virtual Brain activities contributed to learning, 50/53 participants responded (2 blank, 1 ``NA''). Students (16) reported benefits of seeing the pathways, structures, and their connections. Some elaborated that this helped them visualize the systems later while studying. In another theme, students (15) mentioned that they were visual learners and/or that the lessons helped them visualize the material. In response to what would make the Virtual Brain activities more useful for learning, 37/53 responded (9 blank, 7 ``NA''/``nothing''). One theme centered on the amount of information with suggestions for more details or examples in and outside the VR lessons (10). Another theme addressed implementation in the classroom, with recommendations for additional time exploring the lessons, more space in the classroom, and greater clarification of the device controls (12).

In summary, student evaluations suggested the Virtual Brain lessons are valuable tools for learning about system pathways in the classroom. Student evaluations also provided useful feedback for improving implementation. 

\section*{General Discussion}
In this study we developed and evaluated guided tours through 3D narrated diagrams of the human brain. The lessons teach learners about functional anatomy in the visual and auditory systems. These lessons can be experienced on different devices, including a desktop PC or a VR head-mounted display (following guidelines for cross-platform access in \textcite{concannon2019}). We tested three hypotheses: (a) participants would learn from lessons presented on both PC and VR devices (pre-test vs. post-test scores), (b) VR would be more effective for achieving content based learning outcomes (i.e., describe key brain regions involved in processing sensory information and the pathways that connect them), and (c) VR would be more effective for achieving experience-based learning outcomes (i.e., enjoyment and ease of use). We assessed content learning using a drawing/labeling task on paper (2D drawing) in Experiment 1 and using a Looking Glass autostereoscopic display (3D drawing) in Experiment 2. 

Supporting our first hypothesis, participants showed significant content-based learning for both devices. Against our second hypothesis, we found no significant differences between PC and VR devices for content-based learning outcomes. This result could not be explained by (mis)alignment of teaching and testing methods as results did not differ when testing was done in 2D (Experiment 1) or 3D (Experiment 2). Supporting our third hypothesis, VR far exceeded PC viewing for achieving experience-based learning outcomes in both experiments. Thus, our UW Virtual Brain Project lessons were effective in teaching functional neuroanatomy. Although knowledge about functional neuroanatomy was similarly accessible across devices, VR was more enjoyable and easier to use. In designing our lessons, we also aimed to prevent motion sickness. Responses to the SSQ for all experiments suggests these efforts were effective, as mean responses were between none and slight for all symptoms, and were never reported as severe. 

To consider why we found no difference between devices for content learning, we return to our discussion of prior work in the introduction. Prior evidence suggests that active exploration and stereopsis improve content-based learning outcomes. Yet, research investigating students' abilities to learn laboratory procedures in narrated lessons reported no differences or worse performance under 3D VR viewing \parencite{makransky2019, makransky2020}. They attributed worse performance to distractions from immersion in VR. In our study, it is possible that distractions from increased immersion within the narrated lesson cancels the benefits of VR for learning 3D structures. It is also possible that VR lessons could have disproportionately benefited students with lower visual-spatial ability \parencite{cui2017, bogomolova2020}. However, because we did not collect measures of visual-spatial ability, we cannot test that possibility with the present data. 

The result that VR and PC viewing are comparable for achieving content-based outcomes may be a positive outcome. It means that learners can have similar access to learning about functional neuroanatomy through multiple platforms. Thus, our Virtual Brain lessons can accommodate learners who  do not have VR headsets or who would have difficulty with VR due to factors like motion sickness or lack of stereoscopic depth perception. 

In this study, we also conducted a classroom implementation, which incorporated the \textit{Virtual Visual System} and \textit{Virtual Auditory System} lessons within a 75-minute lecture of an undergraduate course \textit{Psychology of Perception}. At the end of the semester we gathered indirect measures on learning. These measures reflected students' self-report on the efficacy of the Virtual Brain lessons. The measures also provided feedback on which aspects of the lessons contributed to learning and which aspects could be improved. Of the three learning outcomes we evaluated, students' ratings indicated that the lessons were most effective for making progress on the outcome we prioritized while designing the lessons: learning system pathways. In free responses, students reported that the lessons were most helpful for seeing the pathways and regions and for visualizing the material. Students indicated the lessons could be improved by including more examples and more time in VR to explore the lessons. Both of these aspects would be especially beneficial to students with relatively low visual-spatial ability.
 
With the rise of portable VR headsets and enthusiasm for integrating technology into classrooms, it is important to consider the potential roles of VR in education. In our view VR provides a lens, similar to a microscope or telescope, for transporting learners to environments they could not otherwise inhabit. But, just as students do not spend entire classes with a microscopes attached to their face, they should not spend an entire class in a VR headset isolated from real-world interactions with their peers and instructor. Some display systems, such as CAVEs, can reduce the isolating aspects of HMDs by allowing multiple simultaneous viewers of the content \parencite{cruz1993surround}, but these display systems are also less portable. A possible best-case scenario would adopt a multi-user, networked virtual environment where content is viewed collaboratively via multiple HMDs, such as in \textcite{fischer2020volumetric}. Still, we believe VR should supplement and support conventional classroom teaching techniques, rather than replace them \parencite{mantovani2003, markowitz2018}.

Our results suggest students are enthusiastic about learning in VR, as supported by the high ratings for VR on the individual items of the experience questionnaire, including wanting to use these kinds of experiences for future studies and recommending these experiences to friends for learning or for fun. Although we did not assess long-term outcomes, these positive experiences in VR have the potential to spark interest and lead to greater future engagement with the material \parencite{hidi2006}. However, it is also possible that greater situational interest may not lead to greater learning in the future \parencite{renninger2015}. To maintain  enthusiasm for VR in the classroom, one challenge will be avoiding ``technological obsolescence'' that can arise from using the same VR lessons while failing to keep pace with rapid advances in VR technology \parencite{vergara2020}. 

The UW Virtual Brain Project is an ongoing effort to create interactive diagrams for teaching Sensation and Perception. We are developing lessons for additional perceptual systems with greater learner interactivity (i.e., learners activating signals that propagate through the systems), and incorporating text caption options to increase accessibility. Our lessons are freely available online for educational use and we post new lessons as they become available https://github.com/SchlossVRL/UW-Virtual-Brain-Project. 

\section*{Acknowledgments}
We thank Shannon Sibrel, Zachary Leggon, Autumn Wickman, Ana Ramos Contreras, Yuke Liang, Marin Murack, Nina Sugaya, Lauren Ciha, Amber Westlund, Brianne Sherman, Lexi Soto, Amanda Zhang, Andrew Liu, and Mohan Ji for their help with this project. This work was supported in part by the Ziegler Foundation, the Office of the Vice Chancellor for Research and Graduate Education at the University of Wisconsin--Madison, the Wisconsin Alumni Research Foundation, and the National Science Foundation (BCS-1945303 to KBS).

\printbibliography


\end{multicols}

\newpage

\section* {Supplemental Material}
This Supplemental Material file includes the following content:\\

\noindent Appendix A. Creating the virtual environments \\
Appendix B. The UW Virtual Brain lessons\\
Appendix C. Testing materials\\
Appendix D. Supplemental data and analyses

\section* {A. Creating the virtual environments}

We created the UW Virtual Brain 3D narrated diagrams using a processing pipeline that converts raw volumetric magnetic resonance imaging (MRI) data into a 3D surface format and then into a Unity game engine format. Our target audience and learning outcomes are distinct from prior work on VR brain models, which primarily focused on surgical training \parencite{larsen2001virtual, 10.3389/fninf.2013.00010, bernardo2017virtual}. However, we build on prior approaches for constructing VR models out of MRI data. Early approaches to reconstructing and rendering 3D brain data adopted direct volumetric rendering of aligned MRI image stacks \parencite{lawonn2018survey} used in neuroanatomy lectures \parencite{Kockro2015}. Recent approaches employ game engine technology such as Unity or Unreal, which are free for non-commercial purposes, and contain editing environments for creating content and compiling applications for multiple forms of hardware (e.g., desktop, mobile VR HMDs). Game engine technologies rely on polygonal mesh formats as the primary rendering primitive rather than volumetric image data. For this reason, more recent work has adopted pipelines that convert brain MRI data to mesh formats \parencite{stepan2017, ekstrand2018immersive}. Although \textcite{stepan2017} supports the Oculus Rift VR HMD, it relies on PrecisionVR's Surgical Theater, proprietary software aimed at medical professionals. \textcite{ekstrand2018immersive} use Unity for their VR environment creation, but focused solely on anatomical structures.

\subsection*{MRI data} We acquired MRI data at the Waisman Center in Madison, WI using a GE Discovery Medical 3T MRI scanner equipped with a 32-channel head coil (GE Healthcare, Inc., Chicago, IL, USA).  First, we acquired a whole-brain structural T1-weighted scan (2.93 ms TE; 6.70 ms TR; 1 mm\textsuperscript{3} isotropic voxels). This anatomical scan was followed by a diffusion-weighted sequence consisting of two opposite phase-encoded 48-direction scans (anterior-posterior [AP] and posterior-anterior [PA] phase-encoding; 6 b0 images per scan; 76.7 ms TE; 8.1 s TR; 2x2x2 mm\textsuperscript{3} isotropic voxels; b=2000 s/mm2; reconstruction matrix FOV: LR 212 mm x AP 212 mm x FH 144 mm). 

\subsection*{3D model extraction} We began by estimating mesh geometry from a whole-brain structural T1-weighted scan. We imported the anatomical scan into FreeSurfer, an open source software suite for processing and analyzing brain imaging data \parencite{fischl2012freesurfer}. Using the input T1 volume, FreeSurfer generates surface-based reconstructions of the brain and identifies and labels various brain structures and regions of interest (ROIs). 


\subsection*{3D model creation} In creating the 3D diagrams, we extracted 3D mesh geometry from the source MRI images similar to approaches in \textcite{zhao2014method}. Using CloudCompare, open-source software for processing point cloud datasets \parencite{girardeau2015cloud}, a Poisson surface reconstruction algorithm was used to generate a 3D surface from FreeSurfer's pial surface estimate \parencite{kazhdan2006poisson}. This resulted in a polygonal 3D model representing the entire brain surface (excluding the cerebellum). For the visual and auditory systems, we identified ROIs using either FreeSurfer's default parcellations or Glasser's HCP-MMP1.0 atlas \parencite{glasser2016multi}. We extracted individual brain regions by segmenting the FreeSurfer or Glasser atlas point data by region, and constructing 3D surface models from the individual regions. Regions were further edited for appearance using mesh smoothing algorithms and additional hand adjustments in Blender \parencite{blender}. 

\subsection*{White matter pathway creation} We used a combination of probabilistic tractography and manual approaches to create visual and auditory white matter pathway models. Large white matter projections were identified probabilistically using the MRtrix2 package \parencite{basser2002diffusion,behrens2003characterization,calamante2010track,conturo1999tracking,mori2002fiber,parker2003framework,tournier2004direct,tournier2007robust,tournier2012mrtrix}. Our diffusion-MRI pre-processing and tractography pipelines have been described previously elsewhere \parencite{miller2019linking,allen2015altered,allen2018retinothalamic}. We probabilistically identified white matter projections for the optic nerves (optic nerve head to optic chiasm), optic tracts (optic chiasm to lateral geniculate nucleus), optic radiations (lateral geniculate nucleus to V1), and acoustic radiations (medial geniculate nucleus to primary auditory cortex). Spurious fibers were manually removed from each pathway. The cleaned fibers were then smoothed in Blender \parencite{blender} before being imported into Unity where they were aligned to the existing brain model. For several smaller pathways where probabilistic tractography was not feasible, the pathways were manually constructed based on gross anatomy. Those pathways which were manually constructed include the auditory pathways connecting the cochlea and cochlear nucleus (vestibulocochlear nerve; CN VIII), cochlear nucleus and superior olive, superior olive and inferior colliculus, and the inferior colliculus and medial geniculate nucleus. All pathways---including those manually generated or derived from tractography---were rendered as tubes within Unity. The manually drawn pathways were created by stretching and scaling existing pathways until they roughly matched the correct shape. Pathways were also edited slightly in some cases to ensure clear cortical connections at their start and end points.

\subsection*{User interaction} We next imported the 3D region models into Unity \parencite{unity}. As the brain surface and brain region models generated from FreeSurfer and CloudCompare were already positioned correctly relative to one another, they could be imported into Unity without requiring any edits to their position or scale. While designing the 3D lessons, we reduced geometric complexity and abstracted elements of the perceptual systems, such as number of fibers depicted, to achieve sufficient frame rates for visually comfortable VR experiences. We carefully considered which elements to reduce or abstract while preserving the overall structure of the perceptual system. For both lessons, the brain is placed floating in the center of a cube shaped room, with blue grid-lines marking the boundaries of the room.

Once the brain region models were in place, the track, stations, and anatomy models were positioned around them. Anatomy models, namely models representing the eyes and ears, were purchased from various 3D asset stores. Using a plugin to view the original MRI volume within the Unity editor, they were scaled and positioned as closely to the eyes and ears in the original MRI scan as possible. We placed information stations at locations where learners would stop to hear the audio narration. We constructed the track manually using three-dimensional splines so that it guides the learners through each information station. We decided to confine motion along a clearly marked yellow track to help viewers anticipate their motion trajectory. Such anticipation can reduce motion sickness symptoms in VR. Learners do not embody an avatar in VR, but a circular yellow platform appears where their feet would be located, with arrows that point forwards along the track. We incorporated the platform after receiving feedback on early prototypes that some learners who were averse to heights felt uncomfortable floating on the track, high above the virtual floor.

The visual properties of the brain are designed to change as the learner experiences the lesson. For example, when learners enter the interior of the brain, the brain surface near the learner fades away so it does not obstruct their view. Certain brain structures also fade as the learner passes through them, as described in the section for each lesson. Text labels rotate side to side so that they always face the learner's head position. Some labels also rotate up and down to a certain extent so they remain easy to read. Labels are also visible through other objects, so they are not obscured by the complex geometry inside the brain.

\section*{B. The UW Virtual Brain lessons} 
In this section we provide the synopsis and script for the \textit{Virtual Visual System} and \textit{Virtual Auditory System}. Each system has six information stations that learners visit as they travel through the virtual brain with audio narration at each station. The numbers 1-6 in the audio scripts refer to each station seen in Figure 1 in the main text. The third script is the narration for the practice experience for both the VR and PC lessons.

\subsection*{Virtual Visual System}
\subsubsection*{Synopsis of the Virtual Visual System.} In the \textit{Virtual Visual System}, learners explore how sensory input, provided by an image on a TV screen, projects onto different parts of the retina and travels along pathways to primary visual cortex (V1). Station 1 is located outside the brain, and the brain is opaque, preventing viewing of the pathways and limiting the number of potentially distracting elements. The narration provides an overview of the functional significance of the visual system and the role it plays in everyday life. As learners move along the track to Station 2, the brain becomes translucent, revealing the pathways and structures inside. Station 2 is located near the sensory stimulus, a TV screen, where the narration describes the nature of the visual input. The spatial arrangement of the visual field is represented by a color coded scene on the TV screen: the left half of the screen is red and the right half is blue, whereas the top half is light and the bottom half is dark. This color coding is maintained on the retinas, within the fibers, and all the way to the end of the pathway, helping learners follow the processing of different parts of the visual field. The track brings learners through the TV screen, in front of the brain, and then in through the left retina, where Station 3 is located. At Station 3, the narration explains that visual information is transduced from the sensory signal into neural signals. Station 4 is located at the optic chiasm. The narration describes the process of crossing over to the opposite (contralateral) side of the brain. The track then crosses the optic chiasm and travels into the right half of the brain, continuing to the lateral geniculate nucleus of the thalamus (LGN). Station 5, near the thalamus, includes narration describing how the signal passes through the LGN and onto the secondary sensory nerve fibers. As learners leave this station, the left (blue) half of the brain structures fade out as they approach the last station. Station 6 ends the track at the right V1 where the red (left) half of the TV image can be seen. The narration describes retinotopic organization and its functional significance. From the final vantage point, the learner can look back and appreciate the full pathway from the TV screen to V1.

The general learning outcome for the lesson is to describe the sensory regions and pathways. This goal includes items such as describing how information crosses from both eyes into separate pathways and listing the key steps/stations of the system. The lesson also briefly touches on outcomes focused on describing sensory input, like identifying the visual field map, and explaining the general purpose of the system. 

\begin{flushleft}

\subsubsection*{Script for narration in the Virtual Visual System}

\textbf{Station 1: Start.} Welcome to the Virtual Brain visual pathway demo.  In this demo you will fly through the brain stopping at a series of stations. You will learn how a visual scene is projected onto the eyes, and then travels along the visual pathway. Many structures in the visual pathway work together to allow us to perceive the people, objects, and events we encounter on a daily basis. Move along the track. As we approach the brain, the surface will become transparent so we can see the structures and pathways inside.\medskip

\textbf{Station 2: TV over shoulder.} Here we are looking at a person's brain, as if we are looking over their shoulder. In front of the brain you can see what that person would be seeing. We have colored the left half of the image on the TV in red, and the right half in blue, so we can show you how different parts of the image travel along different tracts in the brain. Similarly, we shaded the top half of the image lighter, and the bottom half darker.\medskip

\textbf{Station 3: Eyes.} Look behind you at the TV and notice how the TV image projects onto each eye. For example, you can see that the red colors project onto the opposite side of each eyeball.  Also notice that the image is flipped upside down with the lighter colors from the upper part of the screen projecting onto the bottom of the eyes. This flipping happens because of the way light passes into the eye. Next, we will follow the visual signal into the eyeball. When information reaches the back of the eye, called the ``retina'', it is converted to electrical signals. Those signals then travel along the optic nerve and into the brain. The fibers you see are approximations of the real fibers, which are actually bundles containing hundreds of thousands of individual axons. Move along the path to follow the signal along the optic nerve through the right eye.\medskip

\textbf{Station 4: Optic chiasm.} Look down and to your right, we have reached a junction known as the Optic Chiasm. Notice that half of the fibers from each eye combine and cross over into only one side of the brain. You can see that the light red and dark red parts of the visual field cross over into the opposite side of the brain. Next, you will follow the fibers of the optic tract. Keep moving yourself along...\medskip

\textbf{Station 5: Thalamus/LGN.} The pathway travels through a relay structure called the Lateral Geniculate Nucleus of the thalamus (or LGN). Take a moment to look back at the flow of information from the TV screen, you can see that the colors from the red side of the image project to the opposite side of the eyeball. You can see from the color of the fibers after the optic chiasm, that information from each side of the visual field has crossed over into opposite hemispheres of the brain. Keep following the track, the next destination of the visual pathway is primary visual cortex (or V1).\medskip

\textbf{Station 6: V1.} V1, also known as primary visual cortex, contains a complete map of the visual field on the surface of the brain. This is a feature known as ``retinotopic'' organization. The valley-like structure in front of you, running all the way across the middle of V1 is called the calcarine sulcus, this divides the upper and lower visual field, notice the light red color from the top of the image is seen on the lower bank of V1.
From V1 forward higher brain regions enable more complex processing such as catching a ball that is thrown to you, or recognizing a friend's face in a crowd. 

\end{flushleft}

\subsection*{Virtual Auditory System}


\subsubsection*{Synopsis of the Virtual Auditory System} 
In the \textit{Virtual Auditory System}, learners explore how sound waves emitted by audio speakers stimulate and are transduced by the cochlea before traveling along pathways terminating in primary auditory cortex (A1). Station 1 begins with a view outside the opaque brain that features ears and is flanked by speakers on both sides. The narration gives an overview of the functional significance of the auditory system and its role in daily life. As learners move along the track to Station 2, the brain becomes translucent. Station 2, located near the speakers, plays narration describing the nature of the input. Graphical depictions of  high and low frequency tones appear in the space between the speakers and ears.  The sound waves are shown in purple (a mixture of red and blue) to represent information processed by both ears, and are coded as light/dark for high/low frequencies, respectively. The track brings learners through the right ear and into the cochlea. Here, at Station 3, narration explains how sensory information is transduced from sound waves into neural signals. The colors are now mapped based on which side of the brain processes the information: blue on the right, red on the left, until the information is merged, where everything becomes purple again. This color-coding scheme is designed to help the learner  track the crossing over and mixing of auditory information from the outside world to opposite sides of the brain. Station 4 is located at the initial nerve pathway from the ear into the brain, near the cochlear nucleus, superior olive, and inferior colliculus. The narration describes the cross-over process to the opposite (contralateral) side of the brain. Station 5 is at the medial geniculate nucleus of the thalamus (MGN). The narration describes how the signal passes through the MGN to secondary sensory nerve fibers. The last station, Station 6 is located at A1. The narration describes tonotopic organization and its functional significance. This region is colored purple to indicate processing of information from both ears, however, the tonotopic arrangement of frequencies in the cochlea is preserved in the lightness gradient projected on the surface of A1. From the final station, the learner can look down to see the entire flow of information from the opposite speaker to the current view of A1.

The learning outcome for the auditory lesson is to describe the sensory regions and pathways. For this lesson, we focus on describing how sounds played independently to the left or right ear combine (i.e., cross-over of information), listing the key steps/stations, and naming the relevant fiber pathways of the system. This lesson also briefly touches on learning outcomes like describing the sensory input and the general purpose of the system.

\subsubsection*{Script for narration of the Virtual Auditory System}

\begin{flushleft}

\textbf{Station 1: Start.} Welcome to the Virtual Brain auditory pathway demo. In this demo you will fly through the brain stopping at a series of stations. You will learn how sounds activate structures in the ears, and then travel along the auditory pathway.  Many structures in the auditory pathway work together to allow us to perceive the speech, music, and environmental sounds we encounter on a daily basis.
Here we are looking at a person's brain while they listen to sounds coming from speakers. Move along the track. As we approach the brain, the surface will become transparent so we can see the structures and pathways inside.\medskip

\textbf{Station 2: Sounds/speakers.} A speaker produces sound through vibrations, which lead to pressure waves in the air. The wave for a guitar chord can be seen next to the speakers. [Complex wave plays]. Everyday sound is a mixture of waves with different frequencies. Complex sounds can be broken down into a combination of simpler waves. [Complex wave splits out into five sinusoids]. 
At many stages of the auditory system, high and low frequency sounds are processed separately. Look over to the brain... to illustrate this separation of frequencies, dark colored fibers indicate processing low frequency sounds and light colored fibers indicate processing high frequency sounds. Sound from the speakers reaches both ears. Shades of red indicate the signal from one ear and shades of blue indicate the signal from the other ear. Next, we will follow the auditory signal into the ear.\medskip

\textbf{Station 3: Ear/Cochlea.} When the signal reaches the spiral structure in the inner ear, called the ``cochlea'', it is converted to electrical signals. Different parts of the cochlea respond to different frequencies of sound. Higher frequencies activate the base of the cochlea, shown in light blue, and lower frequencies activate the apex, shown in dark blue.  Those signals then travel along the auditory nerve and into the brain stem. The fibers you see are approximations of the real fibers, which are actually bundles containing hundreds of thousands of individual axons. Move along the path to follow the signal into the brain stem.\medskip

\textbf{Station 4: Brainstem structures.} Look ahead, notice that the signals from each ear partially cross over into the opposite side of the brain. The fibers from the cochlear nucleus remain on the same side. Then the fibers from both cochlea nuclei connect to both superior olives. This comparison of tiny differences in signals from the two ears is used to detect the direction of sounds.
The signal arrives at structures called inferior colliculi, by this point, most of the information comes from the opposite ear. The left inferior colliculus responds mostly to signals from the right ear, and vice versa. Next the signal from both ears is combined as indicated by the purple color of the fibers. Light fibers still indicate signals from high frequencies, and dark fibers still indicate signals from low frequencies. Move along the track.\medskip

\textbf{Station 5: MGN/Thalamus.} The pathway travels through a relay structure called the Medial Geniculate Nucleus of the thalamus (or MGN). The separated mapping of low and high frequencies is preserved. Keep following the track, the next destination of the auditory pathway, is primary auditory cortex (or A1).\medskip

\textbf{Station 6: A1.} A1, also knows and primary auditory cortex continues to maintain the mapping of frequency on the surface of the brain. This is a feature known as tonotopic organization. Notice that light purple color is shown on the front of A1. From A1 forward many of the features of sound important to humans are processed. Higher brain regions enable more complex processing such as the appreciation of music or recognizing a friend's voice in a crowded room.

\subsection*{Virtual practice experiences}
\subsubsection*{Script for narration of the VR virtual practice experience}
\textbf{Introduction.} Welcome to a visualization of the human Brain. You will have the opportunity to explore the brain and to learn about its structure and function. 

\textbf{Calibration.} First, let's make sure you are comfortable with the VR headset. 

Look at the two eye charts in front of you. You should be able to clearly distinguish the letters directly above the red line in both the near and far charts. If the image appears blurry, the headset will need adjusting.

You can adjust the position of the headset on your face, and adjust the straps so that it stays in place and is comfortable. You can also adjust the slider on the bottom right of your headset. With your thumb, move the slider back and forth until you feel viewing is most comfortable and the image is clear. Let the experimenter know when you are done. 

[The experimenter now makes the brain appear by pressing SPACEBAR]
  
\textbf{Start.} Let's go over how to use the controller. To move along the track, use the joystick under your thumb. Push it forward to move forwards, and pull it backward to move backwards. You will keep moving as long as you are pushing or pulling the joystick. You can also move your head and body to look around in any direction.

Look straight in front of you, there is the first of a series of information stations. When you arrive at each one, my voice will play to provide information relevant to that station. Please move toward the first station.

\textbf{Station 1: Outside brain.} When you arrive at a station, rails will appear around you while my voice is playing. Once the rails disappear you can continue moving. Please move toward the next station.  

\textbf{Station 2: Outside brain.} As you approach the brain, the surface will become transparent so that you can see inside. Follow the track into the brain to activate the next station. 

\textbf{Station 3: Inside brain.} Now, take moment to explore the environment. You can practice looking around and moving back and forth along the track. Let the experimenter know when you feel comfortable moving around the space, and we will move on to the lesson plan.

\subsubsection*{Script for narration of the PC virtual practice experience}
\textbf{Introduction.} Welcome to a visualization of the human Brain. You will have the opportunity to explore the brain and to learn about its structure and function. 

[The experimenter now makes the brain appear by pressing SPACEBAR]
  
\textbf{Start.} Let's go over how to use the mouse. To move along the track, use the left and right mouse buttons. Hold down the left mouse button to move forward, and the right mouse button to move backwards. You will keep moving as long as you hold the button down. You can also move the mouse to look around in any direction. 

Look straight in front of you, there is the first of a series of information stations. When you arrive at each one, my voice will play to provide information relevant to that station. Please move toward the first station.

\textbf{Station 1: Outside brain.} When you arrive at a station, rails will appear around you while my voice is playing. Once the rails disappear you can continue moving. Please move toward the next station.  

\textbf{Station 2: Outside brain.} As you approach the brain, the surface will become transparent so that you can see inside. Follow the track into the brain to activate the next station. 

\textbf{Station 3: Inside brain.} Now, take moment to explore the environment. You can practice looking around and moving back and forth along the track. Let the experimenter know when you feel comfortable moving around the space, and we will move on to the lesson plan.

\section*{C. Testing Materials}
\subsection*{Paper drawing/labeling test}
This section includes the full set of instructions and test materials for the paper tests used in Experiment 1. Figure C\ref{fig:GeneralPaperInstructions} shows the general instructions, which apply for both the visual and auditory system tests. Figures C\ref{fig:VisualPaperInstructions} and C\ref{fig:AuditoryPaperInstructions} are the instructions and questions for the visual and auditory system tests, respectively. Participants respond to these questions by drawing/labeling on Figures C\ref{fig:VisualPaperTest} and C\ref{fig:AuditoryPaperTest} which each feature images of the visual and auditory systems, respectively.

\subsection*{Looking Glass drawing/labeling test}
The Looking Glass drawing/labeling tests paralleled those of the paper tests, including the test questions (Figures
C\ref{fig:GeneralPaperInstructions}, C\ref{fig:VisualPaperInstructions}, and C\ref{fig:AuditoryPaperInstructions}). In these tests, participants had four main kinds of tasks across the same five questions: filling in structures/regions with solid color, applying labels to structures, drawing pathways between structures, and painting colors onto structures/regions. See Figures C\ref{fig:VisualLGScreens} and C\ref{fig:AuditoryLGScreens} for screenshots of the touchscreen interface and the Looking Glass display for the five questions in the visual and auditory lesson tests, respectively. 
See Figure C\ref{fig:LGExamples}A for larger example images of the Looking Glass display, including the Leap Motion hand tracking controller. 

For filling in structures/regions with solid color, participants selected a swatch of color from the touchscreen with one hand, which would create a sphere of that color on the tip of the index finger of the 3D other hand. Participants would move their hand to align the 3D hand over the desired structure or region they wished to color. When the 3D hand was hovering over the structure, the structure glowed in the color that was selected. Participants could then press the ``APPLY'' button on the touchscreen, which would apply the color onto the entire structure or selected region. Labeling tasks prompted participants to select a structure label, which would then appear on the 3D hand's index finger. Participants would apply the label when hovering over the desired line/structure, glowing white to show which structure was selected. To draw pathways, participants first select a color swatch, creating the sphere. Upon applying the sphere to the structure, a 3D line would appear, connecting the selected structured to the 3D hand. Participants move their hand to the desired second structure and press apply, which would drop the end of the line onto the structure, connecting the two. Lastly, for painting questions, rather than applying the color to the entire structure, once participants selected the color swatch, participants could move their index finger across the glowing, paintable structures, creating streaks of paint on that structure. 
 
Participants could also rotate the view of the structures in the Looking Glass, undo previous actions, and restart questions, removing all responses in the current view of the brain.

\subsection*{Looking Glass drawing/labeling training}
In Experiment 2, prior to the first pre-test, all participants completed training on how to use the Looking Glass and accompanying touchscreen monitor. Figures C\ref{fig:LGTrainingInst1} and C\ref{fig:LGTrainingInst2} shows the instructions for the Looking Glass training procedure. The entire instructions were read aloud to the participants. Text that appeared on the Touchscreen monitor is colored red. The training used the context of building and decorating a 3D house to teach participants all of the different kinds of interactions (i.e., coloring, labeling, drawing fibers, etc.) that would be needed for testing, as described above. This procedure took approximately 10-15 minutes.

\subsection*{Scoring rubrics}
The paper and Looking Glass drawing/labeling tests (figures C\ref{fig:VisualPaperTest} and C\ref{fig:AuditoryPaperTest}) were graded by two coders using an 18-item rubric. Each item was worth one point. Items 1-14 referred to responses to questions 1-4 on image 1. Items 15-18 referred to question 5, on image 2. The rubrics, with additional details on how to score each item, can be found at https://github.com/SchlossVRL/UW-Virtual-Brain-Project. 

\subsubsection*{Visual system grading rubric}
Items with an asterisk were graded relative to responses to items 1 and 2.\\ 
(1) Red on left visual field. Blue on right visual field.\\
(2) Light on top visual field. Dark on bottom visual field.\\
(3) Flip left visual field onto right of eye and right onto left*.\\
(4) Flip top visual field onto bottom and bottom on top*.\\
(5) Entire field is represented in both eyes (2 hues x 2 lightness).\\
(6) Left and right structure labels match and are correct (each 1/4 point).\\
(7) Two hues (pathways) form each eye.\\
(8) Two lightness levels (pathways) from each eye.\\
(9) Pathways crossover once and only once.\\
(10) A crossover occurs at the optic chiasm.\\
(11) Only one hue per hemisphere at V1.\\
(12) Two lightness levels per hemisphere at V1.\\
(13) Hue from left visual field ends on right V1 and hue from right ends on left V1*.\\
(14) All four stations visited in correct order.\\
(15) Only used one hue in V1.\\
(16) Light and Dark both used in V1.\\
(17) Light and Dark location correct in V1*.\\
(18) Hue used in V1 is correct (hue of left visual field)*.

\subsubsection*{Auditory system grading rubric}
(1) Used purple for frequencies of soundwaves.\\
(2) High frequency soundwaves are light and low frequency are dark.\\
(3) Light and dark of SAME hue used within a cochlea.\\
(4) Light and dark used for inner and outer portions of cochlea.\\
(5) High frequency is light (outer) and low frequency is dark (inner) in cochlea.\\
(6) Left and right structure labels match and are correct (each 1/6 point).\\
(7) Light and dark of SAME hue (pathway) leave matching cochlea.\\
(8) Cochlear Nucleus to Inferior colliculi is a straight pathway with same hues.\\
(9) Some hues cross and some go straight at crossover location.\\
(10) Pathway crossover occurs once and only once.\\
(11) A crossover occurs between cochlear nuclei and superior olives.\\
(12) Starting at and everything after inferior colliculus is purple and stays purple.\\
(13) Light and dark present at A1.\\
(14) All stations visited in correct order.\\
(15) Only used one hue in A1.\\
(16) Light and dark both used in A1.\\
(17) Light and dark splitting left/right (horizontally) in A1.\\
(18) Only used purple in A1. 

\end{flushleft}

\newpage
\setcounter{figure}{0} 
\renewcommand{\figurename}{Figure C}

\makeatletter
\def\fnum@figure{\figurename\thefigure}
\makeatother

\begin{figure}[ht!]
  \includegraphics[width=0.95\columnwidth]{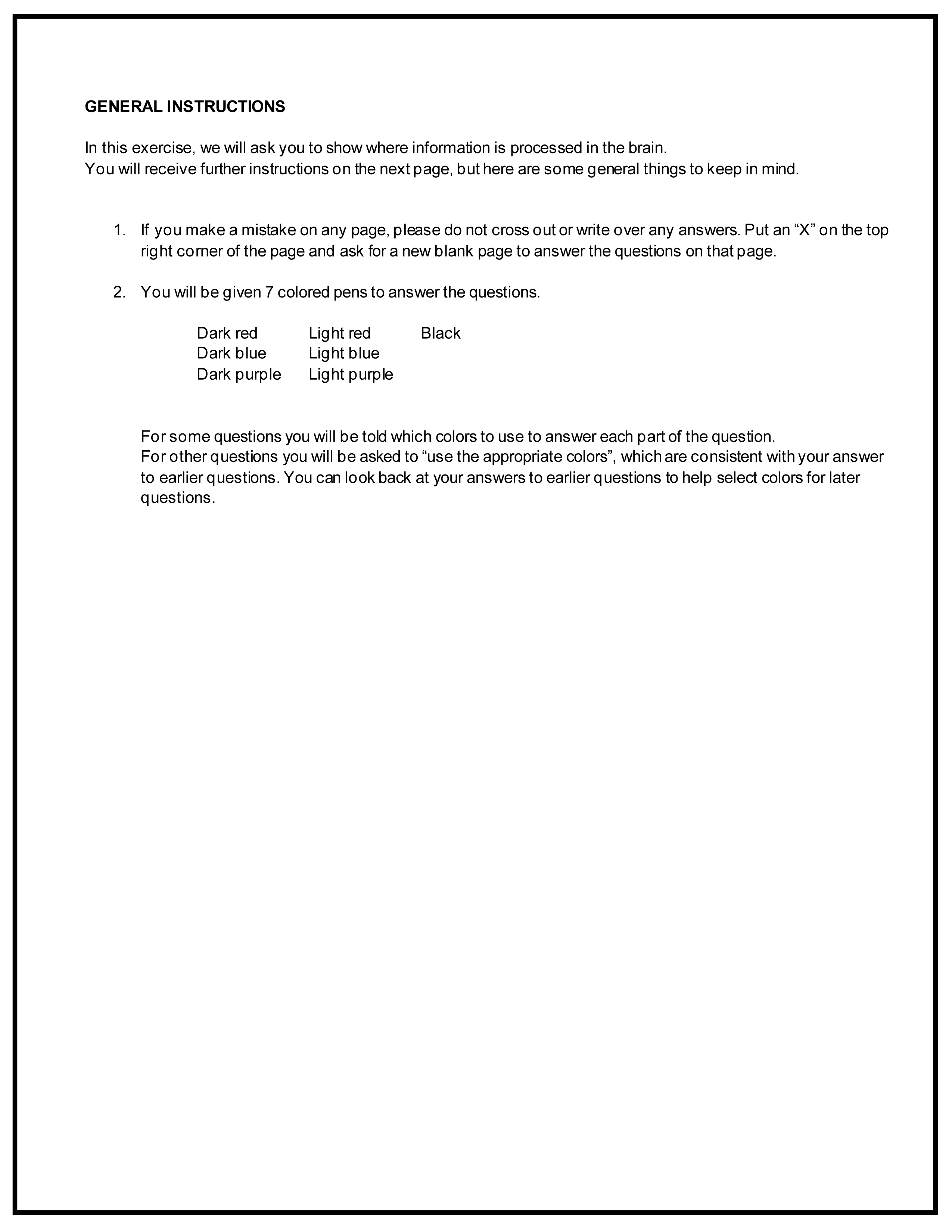}
  \caption{General instructions for the paper tests used in Experiment 1.
  }
  \label{fig:GeneralPaperInstructions}
\end{figure}

\newpage

\begin{figure}[ht!]
  \includegraphics[width=0.95\columnwidth]{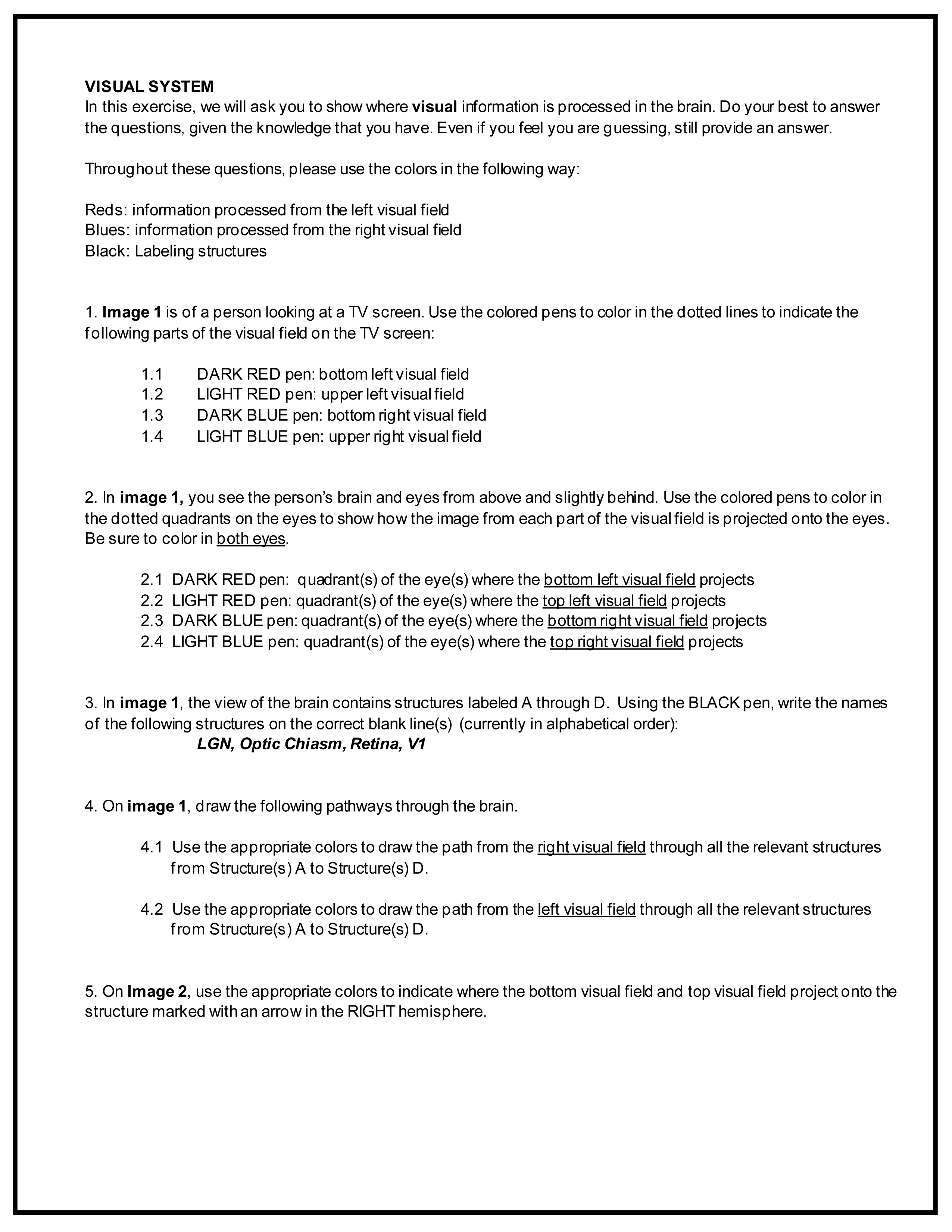}
  \caption{Instructions for the visual system paper test used in Experiment 1.
  }
  \label{fig:VisualPaperInstructions}
\end{figure}

\newpage

\begin{figure}[ht!]
  \includegraphics[width=0.95\columnwidth]{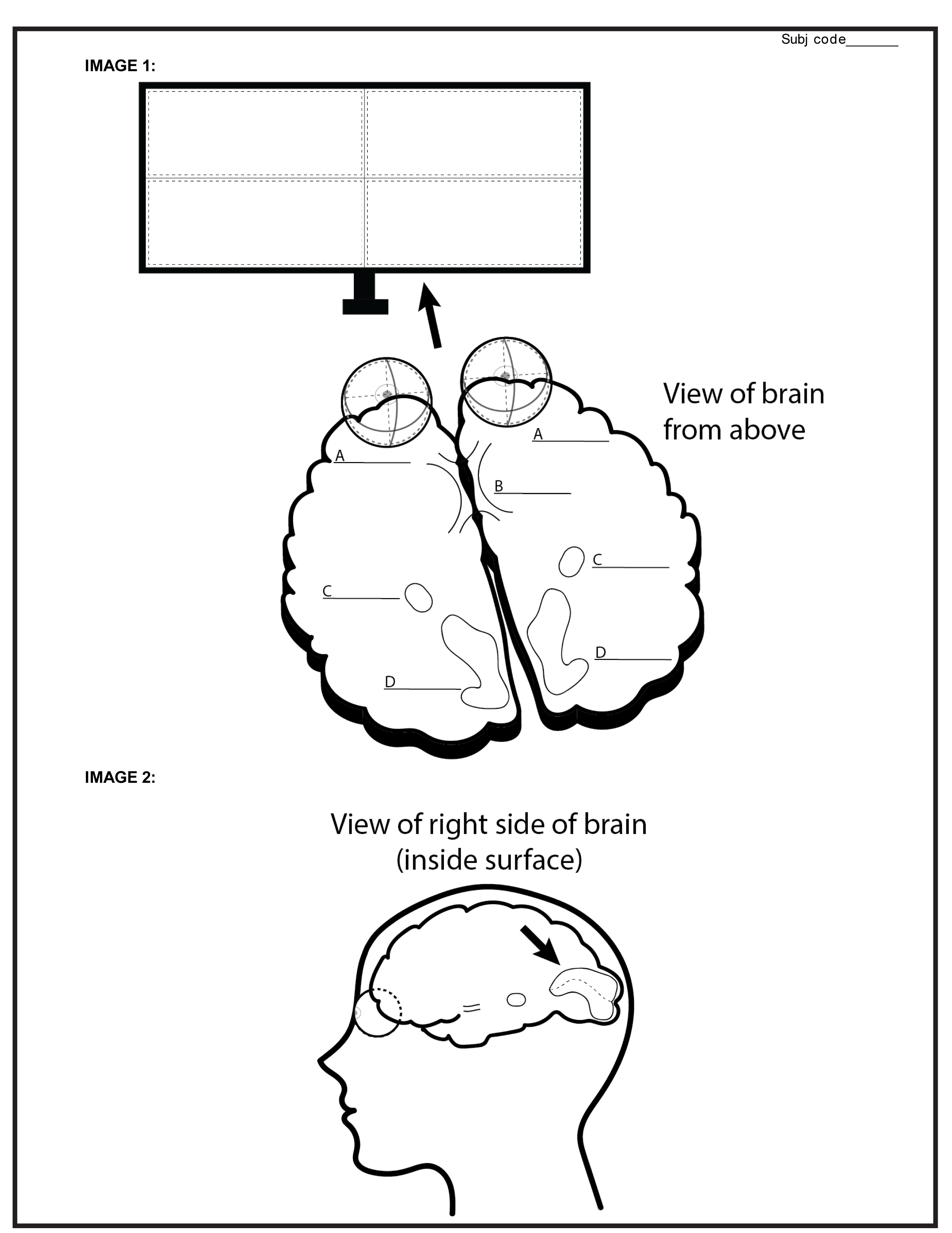}
  \caption{Paper test for the visual system used in Experiment 1.
  }
  \label{fig:VisualPaperTest}
\end{figure}

\newpage

\begin{figure}[ht!]
  \includegraphics[width=0.95\columnwidth]{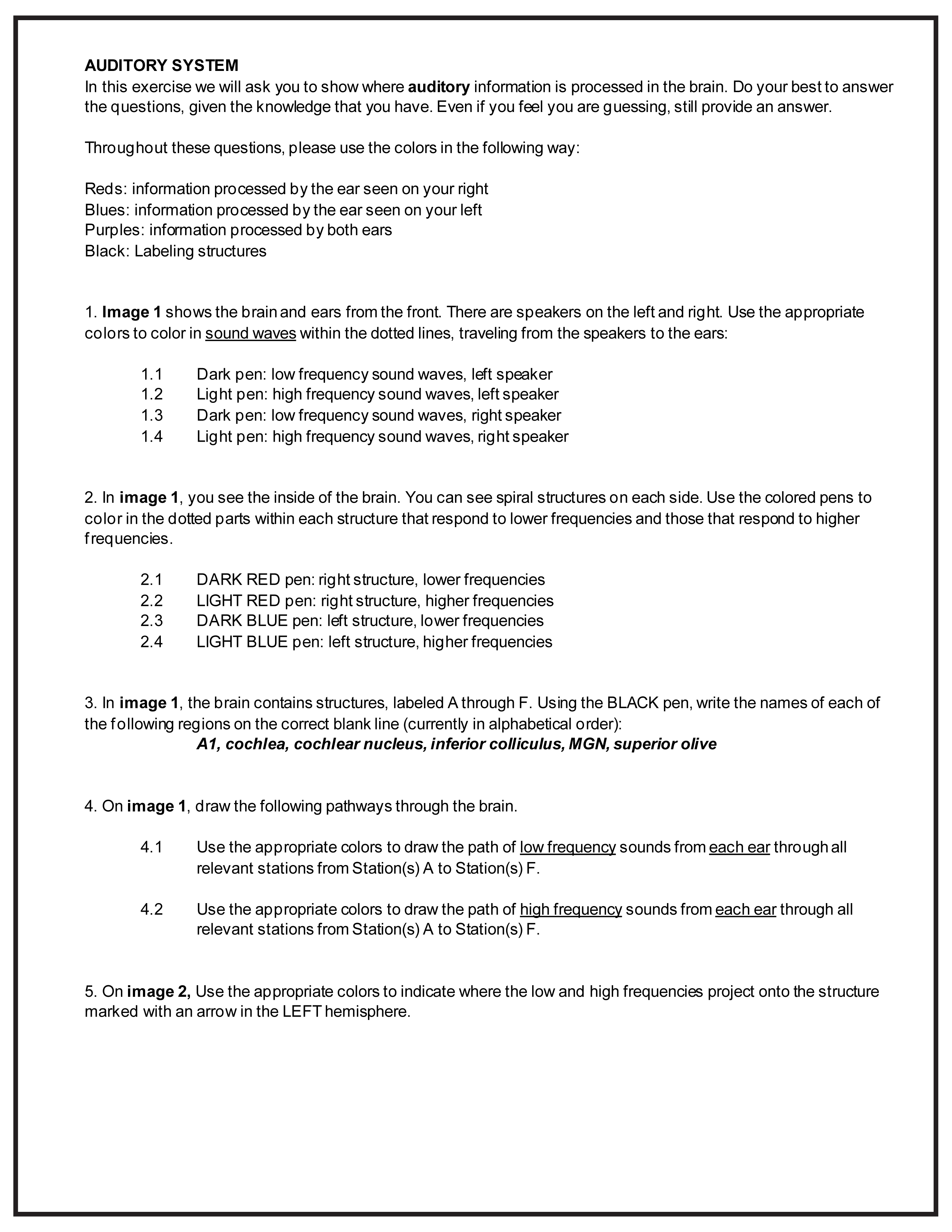}
  \caption{Instructions for the auditory system paper test used in Experiment 1.
  }
  \label{fig:AuditoryPaperInstructions}
\end{figure}

\newpage

\begin{figure}[ht!]
  \includegraphics[width=0.95\columnwidth]{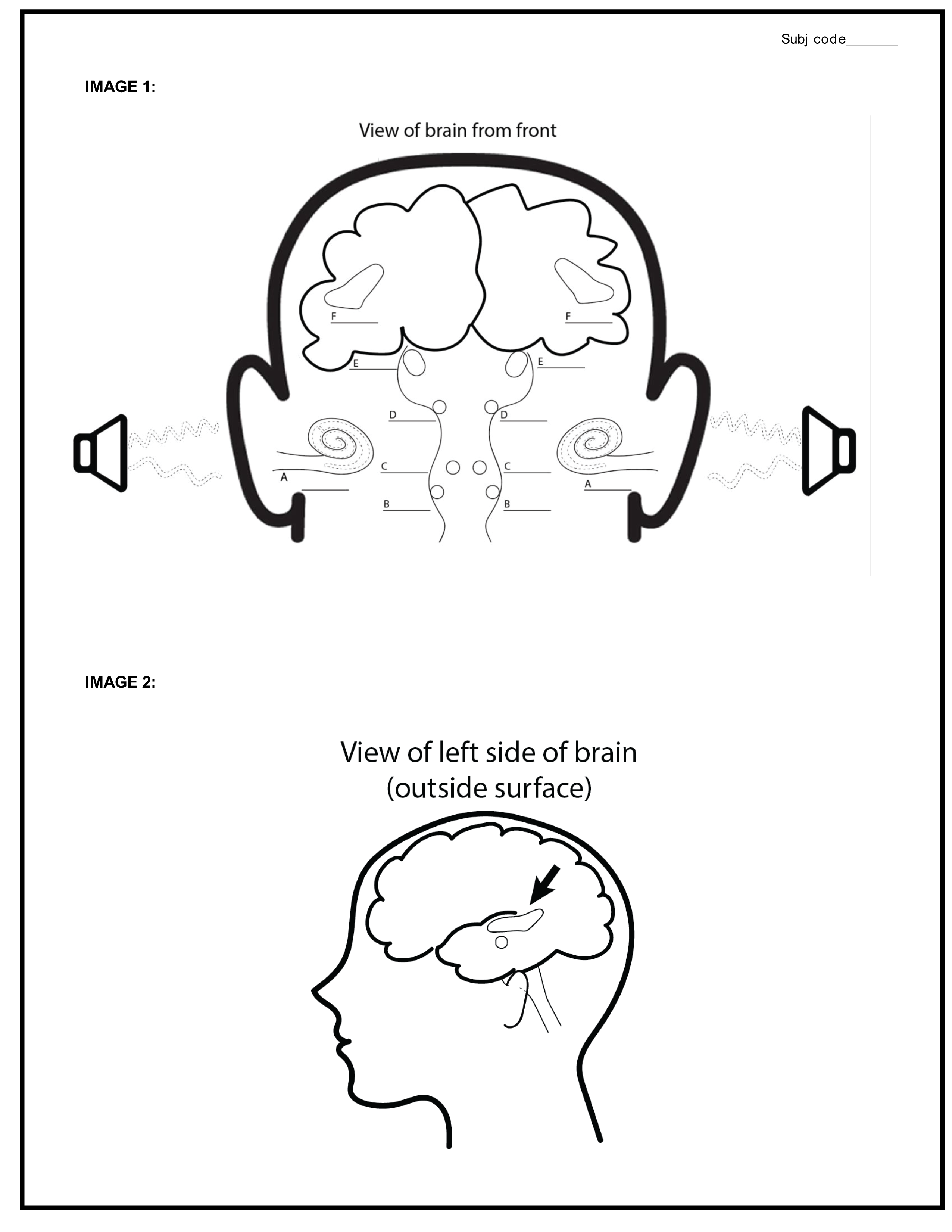}
  \caption{Paper test for the auditory system used in Experiment 1.
  }
  \label{fig:AuditoryPaperTest}
\end{figure}

\newpage

\begin{figure}[ht!]
  \includegraphics[width=0.95\columnwidth]{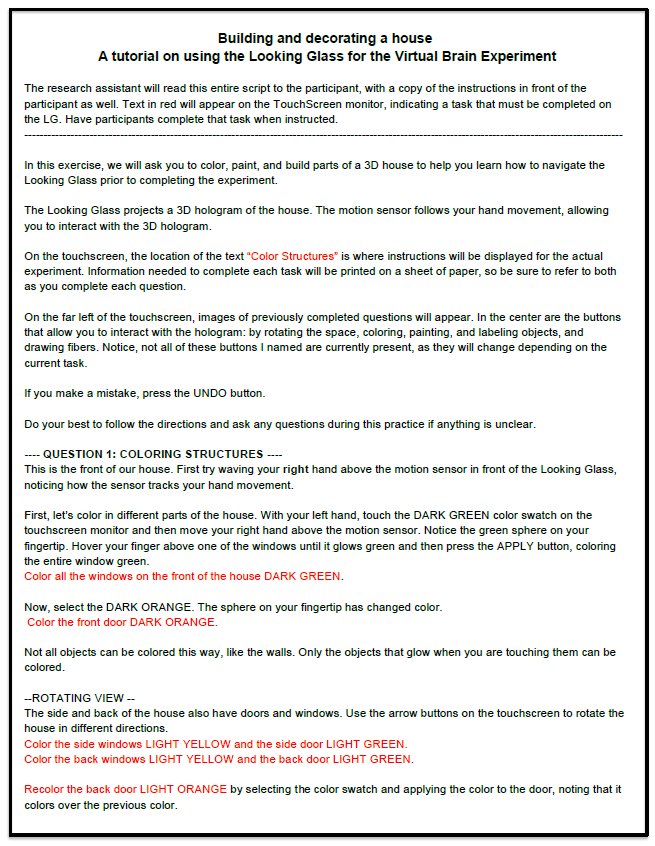}
  \caption{First page of instructions for the Looking Glass training procedure used in Experiment 2.
  }
  \label{fig:LGTrainingInst1}
\end{figure}

\newpage

\begin{figure}[ht!]

  \includegraphics[width=0.95\columnwidth]{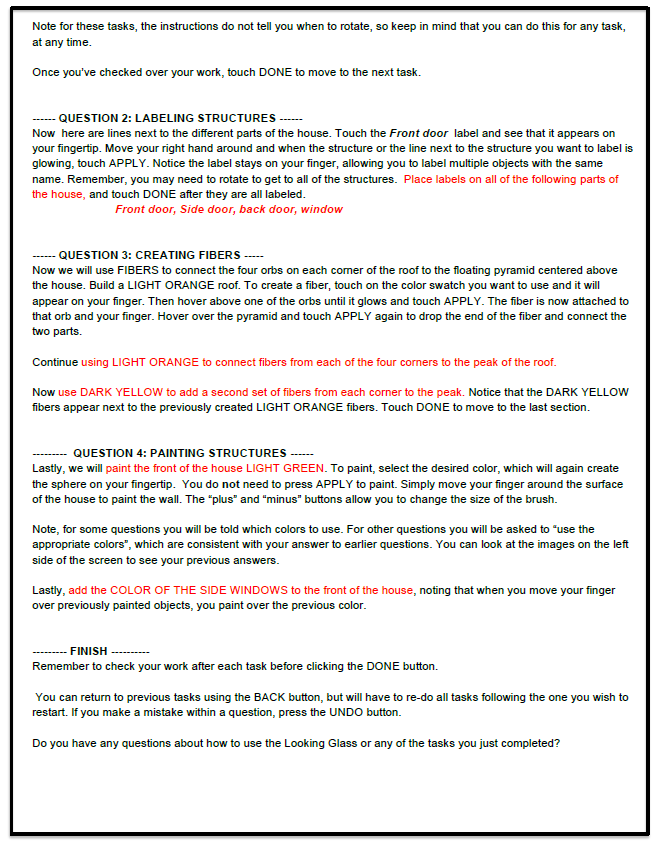}
  \caption{Second page of instructions for the Looking Glass training procedure used in Experiment 2.
  }
  \label{fig:LGTrainingInst2}
\end{figure}

\newpage

\begin{figure}[ht!]
\centering
  \includegraphics{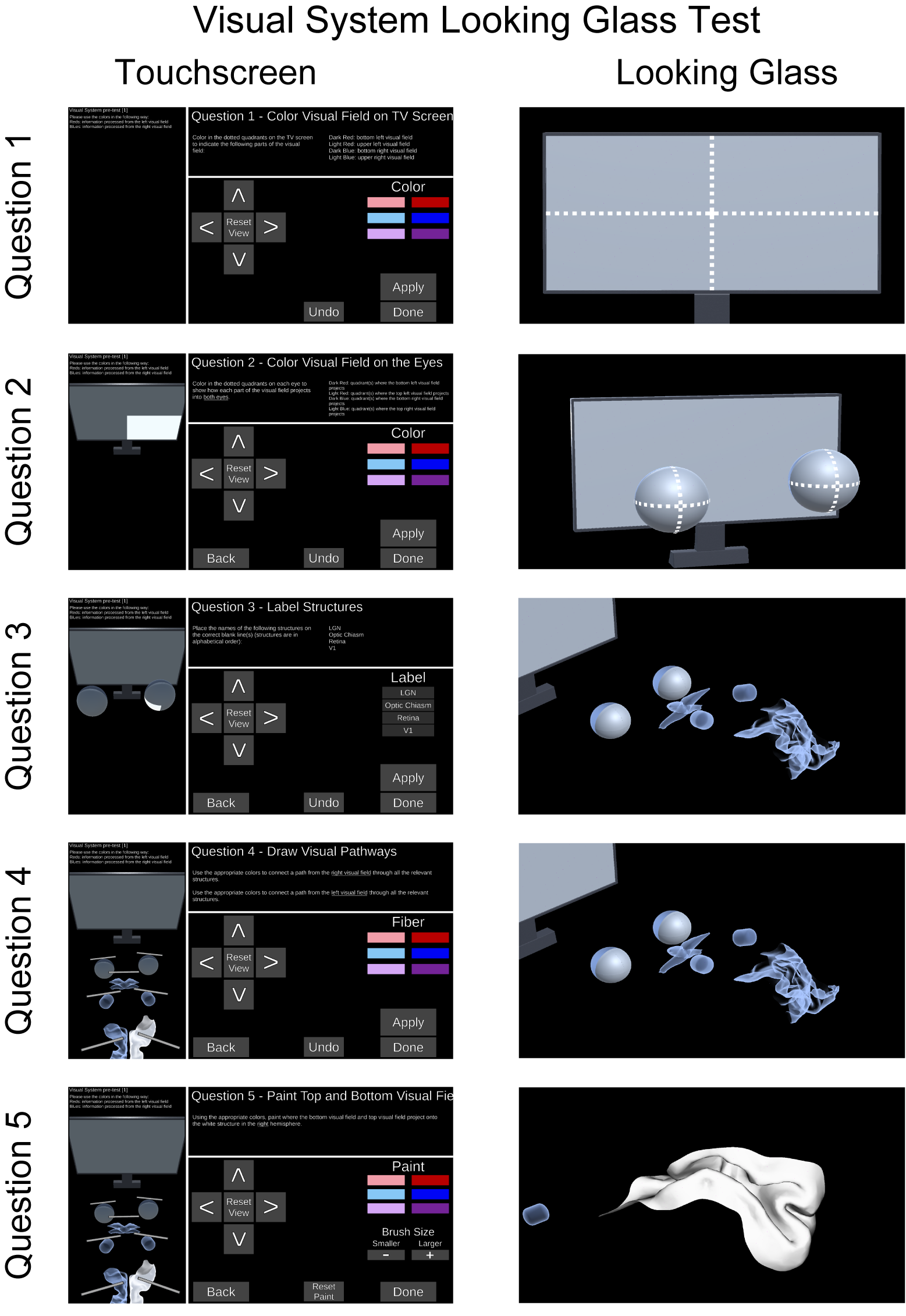}
  \caption{Screenshots of the touchscreen (left) and model views used as input to the Looking Glass (right) for the visual system test in Experiment 2.
  }
  \label{fig:VisualLGScreens}
\end{figure}

\newpage

\begin{figure}[ht!]
\centering
  \includegraphics{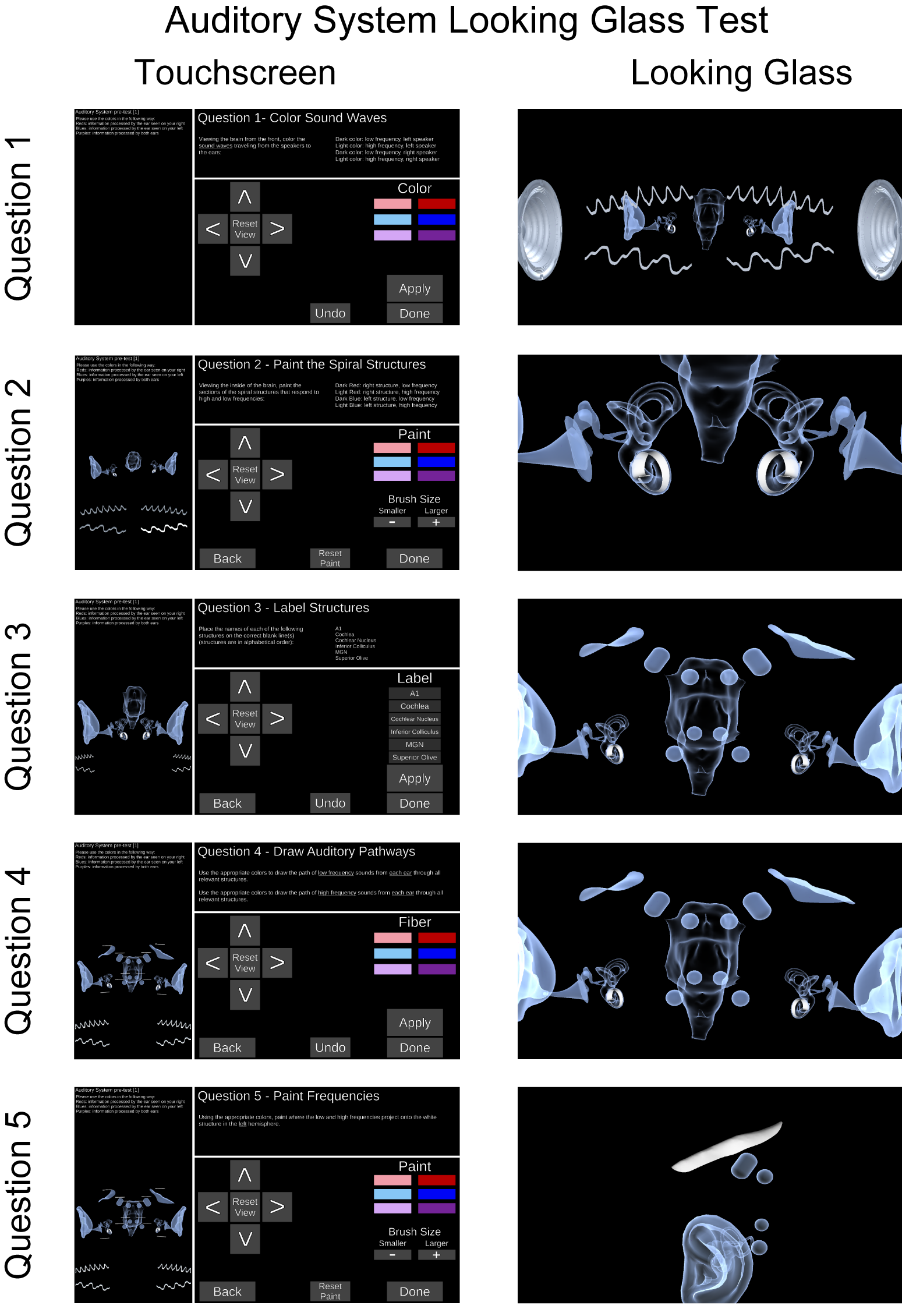}
  \caption{Screenshots of the touchscreen (left) and model views used as input to the Looking Glass (right) for the auditory system test in Experiment 2.
  }
  \label{fig:AuditoryLGScreens}
\end{figure}

\newpage

\begin{figure}[ht!]
\centering
  \includegraphics{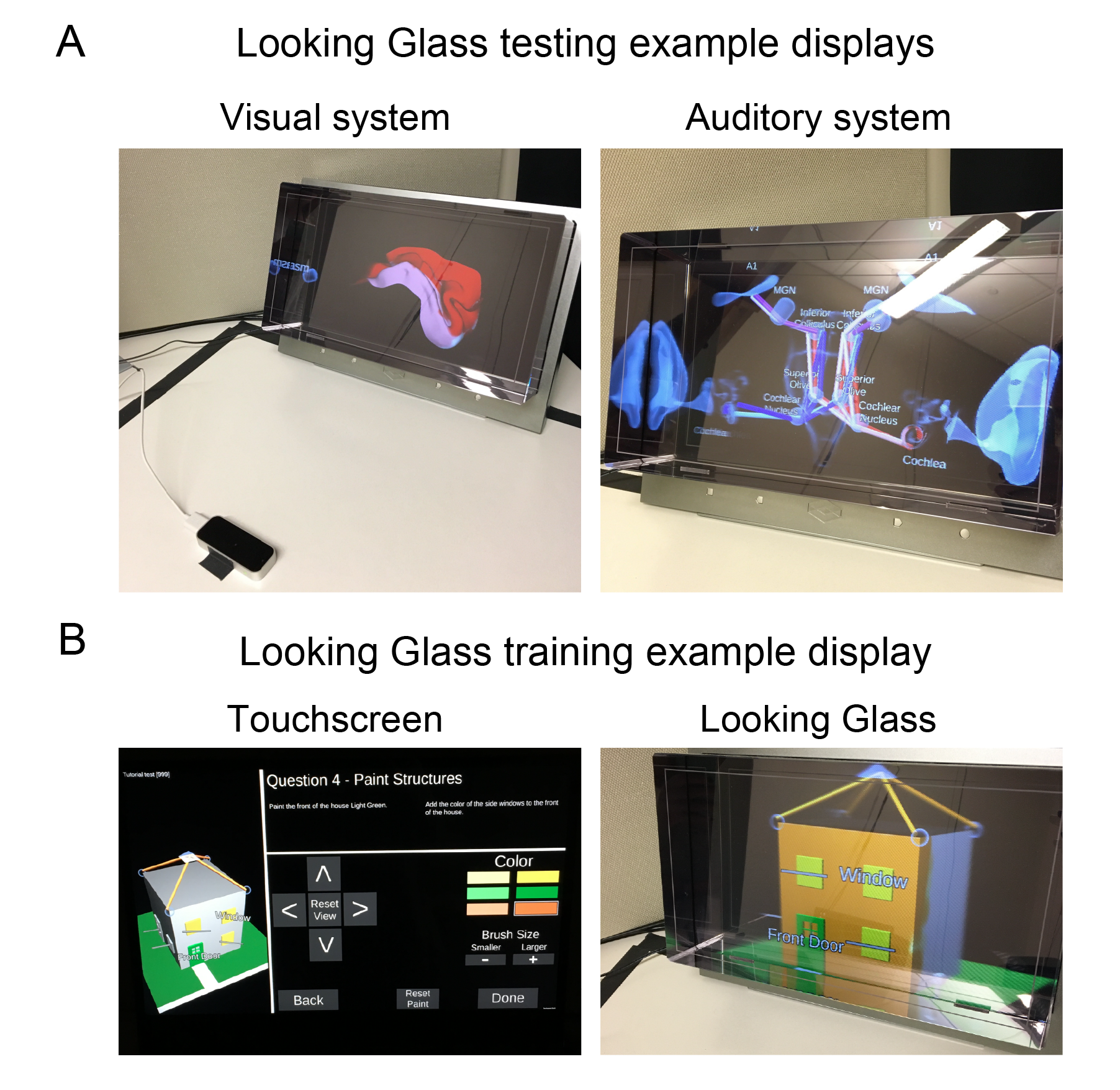}
  \caption{(A) Example Looking Glass displays for the visual system test, in which V1 has been painted, and auditory system test, in which the structures have been labeled and fibers have been drawn. In the lower left of the visual system image, the Leap Motion hand tracking controller is visible. Black tape marked the region the Leap Motion could detect a hand. (B) Example touchscreen and Looking Glass training displays. The 3D house has been painted orange. The doors and windows have been labeled and filled in with green and yellow, respectively and orange fibers connect the four corners to the peak, forming a roof.
  }
  \label{fig:LGExamples}
\end{figure}

\newpage

\section*{D. Supplemental data and analyses} 

\subsection*{Relations between initial test performance and change in performance}

Given previous evidence for relations between between initial scores (baseline) and amount of learning in VR \parencite{zinchenko2020}, we correlated pretest performance and change in performance for each device. In Exp. 1A, correlations were negative, but not significant (VR: $r = -.24, p = .06$, PC: $r = -.24, p = .06$). In Exp. 1B, correlations were also negative (VR: $r = -0.065, p = .481$, PC: $r = -.28, p = .002$), but only PC was significant. In Exp. 2 correlations were again negative (VR: $r = -.261, p = .07$; PC: $r = -.257, p = .07$) but not significant. Taken together, these results suggest a weak, and mostly non-significant trend in which learners with poorer performance initially had a slight tendency to benefit more from the lessons.
 
\subsection*{Results for experience questionnaire items.}
As described in the main text, mean ratings for items 1-6 on the experience questionnaire were highly correlated with each other (see Table D\ref{table:CorMatrix}) so we reduced the dimensions using Principle Components Analysis.   

\subsection*{Results from Simulator Sickness Questionnaire (SSQ)} Figure D\ref{fig:STEMSSQBars} shows mean responses to the four simulator sickness symptoms (scored such that ``none''  = 1, ``slight'' = 2, ``moderate'' = 3, ``severe'' = 4). In all experiments, mean responses to all four symptoms fell between none and slight at all time points. No participants ever reported experiencing severe symptoms. This figure omits date from one participant (Exp. 1A) who was excluded because two SSQ tests were marked as obtained at time point 2. 

\setcounter{figure}{0} 
\renewcommand{\figurename}{Figure D}
\makeatletter
\def\fnum@figure{\figurename\thefigure}
\makeatother

\begin{figure}[hb!]
  \includegraphics[width=0.95\linewidth]{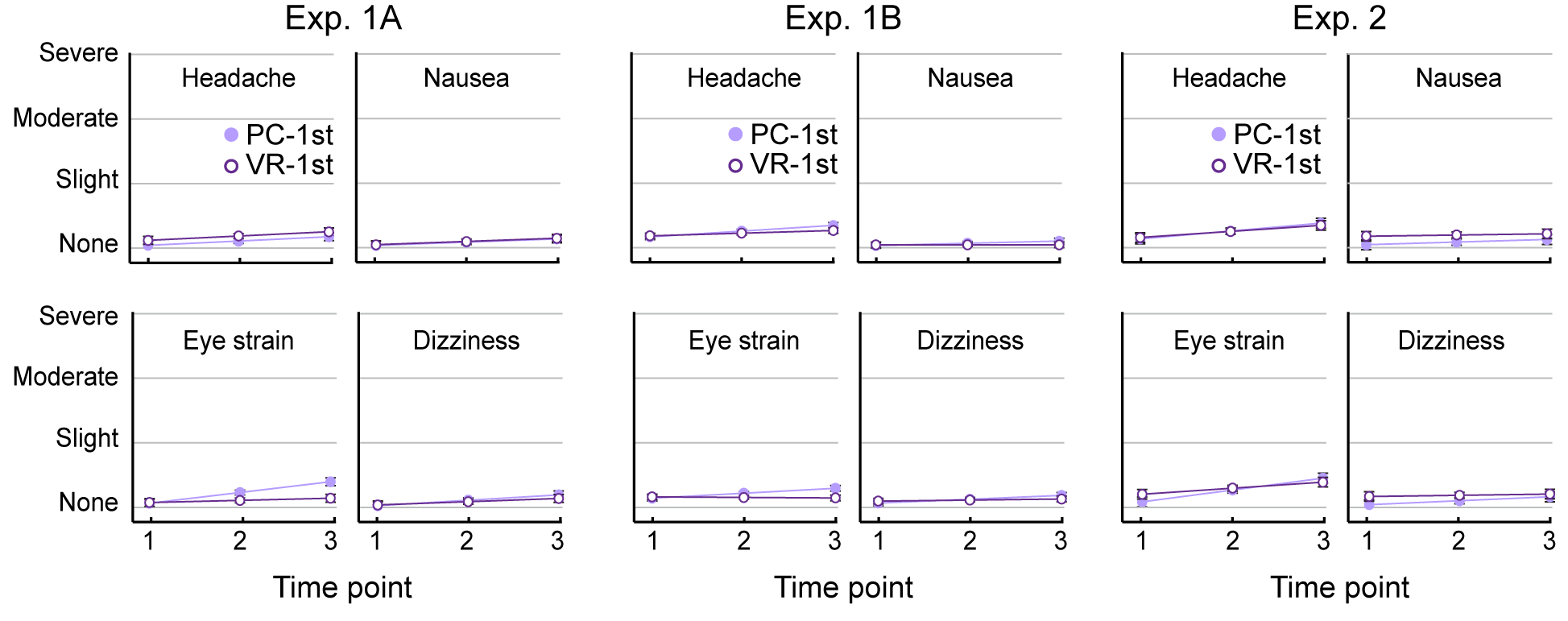}
 \caption{Mean scores on each SSQ symptom for Exp. 1A, Exp. 1B, and Exp. 2 for participants who completed VR-first (dark lines) and PC-first (light lines). Error bars represent SEMs.}
  \label{fig:STEMSSQBars}
\end{figure}

\renewcommand\tablename{Table D}
\makeatletter
\def\fnum@table{\tablename\thetable}
\makeatother

\begin{table}[ht!]
\centering
\begin{tabular}{ lccccccc } 
\toprule
 &  Awe &  Aesthetic & Enjoy & OwnStudy & RecLearn & RecFun & EaseUse\\ 
\midrule
  1. Awe & 1 &  &  &  &  &  &  \\ 

  2. Aesthetic & .81 & 1 &  &  &  & & \\ 

  3. Enjoy & .84 & .83 & 1 &  &  & &  \\ 

  4. OwnStudy & .72 & .70 & .80 & 1 &  &  &  \\ 

 5. RecLearn & .68 & .71 & .77 & .82 & 1 & &  \\ 

 6. RecFun & .82 & .71 & .77 & .72 & .69 & 1 &  \\ 

 7. EaseUse & .24 & .24 & .22 & .17 & .21 & .27  & 1\\ 
\bottomrule
 \end{tabular}
\caption{Correlations of ratings on the 7 items from the Experience questionnaire (Exp. 1A).}
\label{table:CorMatrix}
\end{table}

\end{document}